\documentclass[prx,aps,twocolumn,superscriptaddress]{revtex4-2}

\usepackage[dvips]{graphicx}

\usepackage{amssymb,amsfonts,amsmath,gensymb,eucal}
\usepackage{color}
\usepackage[normalem]{ulem}
\usepackage{natbib}
\usepackage[hidelinks]{hyperref}

\newcommand{\vecp}[1]{\textbf{#1}}

\newcommand{\fext}{\vecp f_{\rm{ext}}}
\newcommand{\fint}{\vecp f_{\rm{int}}}
\newcommand{\Fint}{\vecp F_{\rm{int}}}
\newcommand{\fad}{\vecp f_{\rm{ad}}}
\newcommand{\fsup}{\vecp f_{\rm{sup}}}
\newcommand{\fsupr}{\vecp f_{{\rm sup},r}}
\newcommand{\fvis}{\vecp f_{\rm{vis}}}
\newcommand{\fid}{\vecp f_{\rm{id}}}
\newcommand{\fstr}{\vecp f_{\rm{str}}}
\newcommand{\vel}{\vecp v}
\newcommand{\rv}{\vecp r}

\newcommand{\fintn}{\vecp f_{\rm{int}}^{\,\star}}
\newcommand{\fextn}{\vecp f_{\rm{ext}}^{\,\star}}
\newcommand{\fadn}{\vecp f_{\rm{ad}}^{\,\star}}
\newcommand{\fsupn}{\vecp f_{\rm{sup}}^{\,\star}}

\newcommand{\fd}{\rv;[\rho,\vel]}

\begin{document}

\title{Neural force functional for non-equilibrium many-body colloidal systems}

\author{Toni Zimmerman} \affiliation{Theoretische Physik II,
Physikalisches Institut, Universit{\"a}t Bayreuth, D-95447 Bayreuth, Germany}

\author{Florian Sammüller} \affiliation{Theoretische Physik II,
Physikalisches Institut, Universit{\"a}t Bayreuth, D-95447 Bayreuth, Germany}
\author{Sophie Hermann} \affiliation{Theoretische Physik II,
Physikalisches Institut, Universit{\"a}t Bayreuth, D-95447 Bayreuth, Germany}
\author{Matthias Schmidt} \affiliation{Theoretische Physik II,
Physikalisches Institut, Universit{\"a}t Bayreuth, D-95447 Bayreuth, Germany}
\author{Daniel de las Heras}
\email{delasheras.daniel@gmail.com}
\homepage{www.danieldelasheras.com}
\affiliation{Theoretische Physik II, Physikalisches
Institut, Universit{\"a}t Bayreuth, D-95447 Bayreuth, Germany}

\begin{abstract}
	We combine power functional theory and machine learning to study non-equilibrium overdamped many-body systems of colloidal particles at the level of one-body fields.
	We first sample in steady state the one-body fields relevant for the dynamics from computer simulations of Brownian particles under the influence of randomly generated external fields.
	A neural network is then trained with this data to represent locally in space the formally exact functional mapping from the one-body density and velocity profiles to the one-body internal force field.
	The trained network is used to analyse the non-equilibrium superadiabatic force field and the transport coefficients such as shear and bulk viscosities.
	Due to the local learning approach, the network can be applied to systems much larger than the original simulation box in which the one-body fields are sampled.
	Complemented with the exact non-equilibrium one-body force balance equation and a continuity equation, the network yields viable predictions of the dynamics in time-dependent situations.
	Even though training is based on steady states only, the predicted dynamics is in good agreement with simulation results.
	A neural dynamical density functional theory can be straightforwardly implemented as a limiting case in which the internal force field is that of an equilibrium system.
        The framework is general and directly applicable to other many-body systems of interacting particles following Brownian dynamics.

\end{abstract}

\date{\today}

\maketitle

\section{Introduction}%

Analyzing how a many-body system reacts to controlled stimuli offers a means to understand its collective behavior emerging from interparticle interactions.
Soft matter systems respond to several types of external fields~\cite{Lwen2001,Erbe2008,Menzel2015} such as e.g.~electric~\cite{Velev2006,Vissers2011}, magnetic~\cite{Tierno2007,Lips2021}, gravitational~\cite{Sullivan2002,Eckert2021,Isele2023}, optical~\cite{Faucheux1995}, and mechanical~\cite{Reichhardt2016,FigueroaMorales2022} fields.
The response of the many-body system to the external perturbation can be highly non-trivial, particularly in non-equilibrium situations.
A detailed case-by-case analysis is often required to rationalize the complex dynamics of the system.

Due to the large number of microscopic degrees of freedom in a many-body system, coarse-graining~\cite{Schilling2022} is necessary in order to define a reduced set of relevant variables that encapsulates the physics of the system.
Averaging over the many-body probability distribution function yields an exact one-body force balance equation in overdamped Brownian~\cite{Schmidt2013} as well as in inertial classical~\cite{Schmidt2018} and quantum~\cite{Schmidt2015} systems.
The one-body fields remain sharply resolved in both space and time~\cite{Schmidt2022}.
The force balance equation combined with a continuity equation describes the microscopic dynamics of the system at the one-body level.

Depending on the underlying particle dynamics, different contributions appear in the force balance equation~\cite{Schmidt2022}.
In overdamped Brownian dynamics, the only unknown contribution is the one-body internal force field, $\fint(\rv,t)$, originated by an average of the interparticle interactions resolved in position, $\rv$, and time, $t$.
The response of the system to an arbitrary external force field can be determined provided that the internal force field is known.

In equilibrium, classical density functional theory (DFT)~\cite{Evans1979} is the reference framework to build approximations for the one-body internal force field. 
DFT establishes a mapping from the equilibrium density profile $\rho(\rv)$ to the internal force field $\fint(\rv;[\rho])$ via the one-body direct correlation functional
\footnote{The one-body direct correlation functional $c_1(\rv;[\rho])$ is related to the excess (over ideal gas) free energy functional $F_{\rm exc}[\rho]$ via $c_1(\rv;[\rho])=-\delta\beta F_{\rm exc}[\rho]/\delta\rho(\rv)$ with $\beta=1/k_BT$.
The internal force field is then related to the one-body direct correlation function via $\fint(\rv;[\rho])=k_BT\nabla c_1(\rv;[\rho])$. Here $k_B$ is the Boltzmann constant and $T$ is (absolute) temperature.}
(we indicate with square brackets the functional dependence of the internal force field on the density profile).
Several works~\cite{Cleaver2014,Lin2019,Lin2020,Cats2021,Morales2023,Sammuller2023,Simon2024,dijkman2024} have demonstrated that machine learning is a reliable technique to construct density functionals.
An important difference among these works is the dataset used to train the machine learning model.
These include pairs of external potentials and density profiles~\cite{Lin2019,Cats2021,Morales2023} as well as radial distribution functions of homogeneous fluids~\cite{dijkman2024}.
Recently, we have shown~\cite{Sammuller2023} that a neural network can be efficiently employed to represent the equilibrium functional mapping from the density profile to the direct correlation functional.
The mapping can then be used to construct a neural functional theory~\cite{Sammuller2023}, which for a hard-sphere system outperforms the most sophisticated analytical density functionals~\cite{HansenGoos2006}.
Two- and three-body correlation functions as well as free energy values are accessible via functional calculus implemented with automatic differentiation and functional line integration.

Power functional theory (PFT) demonstrates that fundamental mappings also exist in overdamped and inertial classical and quantum many-body systems in non-equilibrium~\cite{Schmidt2022}.
For overdamped Brownian systems, there is a formally exact kinematic mapping from both the one-body density $\rho(\rv,t)$ and velocity $\vel(\rv,t)$ profiles to the internal force field $\fint(\rv,t;[\rho,\vel])$~\cite{Schmidt2013}.
The functional dependence of $\fint$ is on the whole time-history of both fields, $\rho(\rv,t)$ and $\vel(\rv,t)$, until the current time $t$.
Previous works have constructed analytical approximations to the kinematic mapping~\cite{delasHeras2018}, revealing how the non-equilibrium internal force field is responsible for notable phenomena in colloidal systems.
These include shear migration~\cite{Stuhlmuller2018,Sammller2023}, the emergence of viscosity and structural forces~\cite{delasHeras2020}, lane formation~\cite{Geigenfeind2020}, the governing mechanisms of the time evolution of the van Hove function~\cite{Treffenstdt2021}, and the mobility induced phase separation~\cite{Hermann2019} and freezing~\cite{Hermann2023} in active Brownian particles.

In a recent perspective about dynamical density functional theory~\cite{delasHeras2023}, we showed that a neural network can accurately represent the functional kinematic mapping from the density and the velocity profiles to the internal force field.
We refer here to the trained network as the neural force functional. 
As a proof of concept, we trained a network using only bulk flows (i.e.~$\nabla\cdot\vel\neq0$ and $\nabla\times\vel=0$) in which the external force points only along one Cartesian direction.
In this type of flow, the non-conservative contribution of the one-body external field is limited to a uniform constant force.
Here, we demonstrate that using data augmentation and a local learning approach~\cite{Sammuller2023}, the mapping can be efficiently learnt in general planar geometry (that is, for flows exhibiting non-vanishing curl and non-vanishing divergence of the velocity field).
Furthermore, we showcase the network's predictive capability in several applications.
We generate the one-body fields directly from particle-based computer simulations of overdamped isotropic colloidal particles in steady state under the influence of randomly generated external force fields.
The particles interact with each other via a Lennard-Jones potential.
We then train a neural network to represent the kinematic mapping. 
The internal force field is learnt locally in space.
As a result, the network can be applied to systems of variable size.
We show several applications that use the neural force functional combined with the exact force balance equation:
(i) splitting and analysing non-equilibrium superadiabatic forces,
(ii) performing inverse design by finding the external force field that generates the desired dynamical response of the many body-system (custom flow~\cite{delasHeras2019}),
and (iii) quantifying non-equilibrium transport coefficients.
Finally, we demonstrate that the network is capable to generalize to systems of any length (beyond the size of the simulation box) and to full non-equilibrium situations (beyond steady-states).

Our neural force functional is computationally efficient and delivers results with precision close to simulation data.
The capability to process systems of size much larger than that of the training data opens a route to study the dynamics of macroscopic systems with microscopic resolution at near simulation quality.

\section{Theory}
\subsection{Force balance equation}
We consider a classical system of $N$ interacting particles suspended in a solvent and following overdamped dynamics.
The equation of motion for the $i$th particle is
\begin{equation}
\gamma\frac{d\rv_i(t)}{dt}={\boldsymbol \eta}_i(t)-\nabla_i u(\rv^N)+\fext(\rv_i,t),\label{eq:mo}
\end{equation}
where $\gamma$ is the friction coefficient against the implicit solvent, $\rv_i$ is the position of the particle,
$\boldsymbol\eta_i(t)$ is a Gaussian random force acting at time $t$ on the $i$th particle,
$\fext$ is an external force field,
$\nabla_i$ indicates the derivative with respect to $\rv_i$,
and
\begin{equation}
	u(\rv^N)=\sum_{i}\sum_{i<j}\phi(r_{ij})
\end{equation}
is the total potential energy of microstate $\rv^N=\rv_1,...,\rv_N$ with $\phi(r_{ij})$ being the interparticle pair potential, $r_{ij}$ being the distance between particles $i$ and $j$, and the first sum runs over all particles.
Although for simplicity only two-body potentials are considered here, PFT~\cite{Schmidt2022} is directly applicable to many-body interparticle potentials (see an example in Ref.~\cite{Sammller2023}).

One-body fields are obtained as averages of microscopic operators~\cite{Schmidt2022}.
For example, the one-body density profile is 
\begin{equation}
	  \rho(\rv,t) =
	\Big\langle\sum_{i} \delta(\rv-\rv_i)  \Big\rangle,
  \label{EQdensityDefinition}
\end{equation}
with $\delta(\cdot)$ being the Dirac distribution. The angular brackets $\langle\cdot\rangle$ denote an average that in non-equilibrium is performed at each time $t$ over an ensemble of systems.
The systems differ in their initial microstate and also in the realization of the random forces, i.e.~$\boldsymbol\eta_i$ in Eq.~\eqref{eq:mo}.
Alternatively, in steady state and in equilibrium, the average can be evaluated over time. 

The one-body fields are related via an exact one-body force balance equation~\cite{Schmidt2013,Schmidt2022}
\begin{equation}
	\gamma \vel(\rv,t)=\fid(\rv,t)+\fint(\rv,t)+\fext(\rv,t),\label{eq:fb}
\end{equation}
with $\vel(\rv,t)$ being the one-body velocity field at position $\rv$ and time $t$.
The first contribution on the right hand-side of Eq.~\eqref{eq:fb} is the thermal diffusive term, which is known exactly:
\begin{equation}
	\fid(\rv,t)=-k_BT\nabla\ln\rho(\rv,t).\label{eq:diff}
\end{equation}
Here $k_B$ is the Boltzmann constant and $T$ is (absolute) temperature.

The internal force field, $\fint(\rv,t)$, follows from the internal force density field $\Fint(\rv,t)=\rho(\rv,t)\fint(\rv,t)$, which is simply
\begin{equation}
	\Fint(\rv,t)=-\Big\langle\sum_{i}\delta(\rv-\rv_i)\nabla_i u(\rv^N) \Big\rangle.
\end{equation}
The last contribution on the right hand-side of Eq.~\eqref{eq:fb} is the external force field, $\fext(\rv,t)$.

For overdamped Brownian particles following the equation of motion~\eqref{eq:mo}, the force balance equation~\eqref{eq:fb} is exact and valid in general non-equilibrium situations.
In non-equilibrium steady state and in equilibrium, the one-body fields become independent of time.
Additionally, in equilibrium, the one-body velocity field vanishes everywhere.
A detailed derivation of the force balance equation can be found e.g.~in Ref.~\cite{Schmidt2022}.

The one-body density profile is linked to the one-body current profile, $\vecp{J}(\rv,t)=\rho(\rv,t)\vel(\rv,t)$, via the continuity equation
\begin{equation}
	\dot{\rho}(\rv,t)=-\nabla\cdot\vecp{J}(\rv,t),\label{eq:continuity}
\end{equation}
where the overdot denotes a partial time derivative.
Hence, from a theoretical point of view, the force balance equation~\eqref{eq:fb} can be used in combination with the continuity equation~\eqref{eq:continuity} to find the time evolution of the one-body fields $\rho(\rv,t)$ and $\vel(\rv,t)$.
However, we first need to approximate the unknown internal force field.

In equilibrium, DFT establishes that the internal force field is a functional of the one-body density profile via
\begin{equation}
	\fint(\rv;[\rho])=-\nabla\frac{\delta F_{\rm exc}[\rho]}{\delta\rho(\rv)}.\label{eq:finteq}
\end{equation}
Here, $F_{\rm exc}[\rho]$ is the excess (over ideal gas) free energy functional.
Hence, knowledge of the equilibrium density profile suffices to determine the internal force field (provided that $F_{\rm exc}[\rho]$ is known).

Dynamical density functional theory (DDFT)~\cite{Evans1979,Umberto1999} (for a recent and exhaustive review see Ref.~\cite{teVrugt2020}) approximates the non-equilibrium internal force field by that of an equilibrium system via Eq.~\eqref{eq:finteq}.
However, there are several concerns that challenge the reliability of this adiabatic approximation~\cite{delasHeras2023}.
A mayor one follows directly from Eq.~\eqref{eq:finteq}: the one-body internal force field in non-equilibrium contains conservative and non-conservative contributions but DDFT predicts always an equilibrium-like force field which is therefore conservative.
The non-conservative contribution of the internal force field can be important.
For example, colloidal migration in shear fields~\cite{FRANK2003,Leighton1987,delasHeras2020} and lane formation in oppositely driven mixtures~\cite{Dzubiella2002,Vissers2011,Geigenfeind2020} are physical phenomena that can be rationalized on the basis of the non-conservative contributions of the non-equilibrium internal force field.

\subsection{Power functional theory}

To overcome the limitations of DDFT we use power functional theory (PFT)~\cite{Schmidt2013,Schmidt2022} which is based on a formally exact variational principle for non-equilibrium many-body systems.
In PFT, a functional (with units of power) is minimized with respect to the current profile (or the velocity profile) at fixed density profile and fixed time.
The functional is constructed such that the associated Euler-Lagrange equation is the exact force balance equation~\eqref{eq:fb}.
The functional depends functionally not only on the density (as it is the case in equilibrium DFT) but also on the velocity profile.
Formally, the dependence is not instantaneous on time but rather on the complete history of both fields.
The original formulation of PFT was done using a functional of $\rho$ and $\vecp{J}$~\cite{Schmidt2013}.
However, working with $\rho$ and $\vel$ turns out to be more convenient to construct analytical approximations, see e.g.~Refs.~\cite{delasHeras2018,delasHeras2020} and Sec.~\ref{sec:transport}.

In PFT the internal force field is split into adiabatic and superadiabatic contributions
\begin{equation}
	\fint(\rv,t;[\rho,\vel])=\fad(\rv,t;[\rho])+\fsup(\rv,t;[\rho,\vel]).\label{eq:fadsup}
\end{equation}
The adiabatic contribution is defined as the internal force field of a virtual equilibrium system with the same density profile as the non-equilibrium system.
Hence, the adiabatic contribution can be formally obtained via Eq.~\eqref{eq:finteq} and it is at each time $t$ a functional of the instantaneous one-body density, $\rho(\rv,t)$, only.
In contrast, the genuine non-equilibrium superadiabatic contribution is given by the functional derivative of the excess power functional $P_t^{\rm exc}[\rho,\vel]$
\begin{equation}
	\fsup(\rv,t;[\rho,\vel])=-\frac{1}{\rho(\rv,t)}\frac{\delta P_t^{\rm exc}[\rho,\vel]}{\delta \vel(\rv,t)}.\label{eq:fsup}
\end{equation}
The superadiabatic force is a functional of both $\rho$ and $\vel$ since it inherits the functional dependencies of the generating functional.
Equation~\eqref{eq:fsup} is formally exact but it requires knowledge of the excess power functional.
It is possible to construct analytical approximations to the excess power functional using an expansion in powers of the velocity field~\cite{delasHeras2018}.
Here, we follow a different approach.
Instead of finding approximations for $P_t^{\rm exc}[\rho,\vel]$, we machine learn the relation $\{\rho,\vel\}\rightarrow\fint$ that maps both the density and the velocity profile to the internal force field.
We learn the complete internal force field, i.e.~both superadiabatic and adiabatic contributions, since the splitting can be done a posteriori as we show in Sec.~\ref{sec:split}.

The first step in any machine learning application is the generation of the training set that we discuss in the following.

\subsection{Simulations and training set}

We employ adaptive Brownian dynamics simulations~\cite{Sammller2021} to integrate the many-body equations of motion~\eqref{eq:mo} over time and to generate the data of the training set.
All terms contributing to the force balance equation~\eqref{eq:fb} can be obtained from many-body computer simulations.
The velocity profile can be sampled either directly via the central difference derivative of the position vector with respect to time or indirectly via the force balance equation.
For details see e.g.~the appendix of Ref.~\cite{delasHeras2019}.
The training set contains the one-body density and the velocity profiles as input fields, and the one-body internal force as the output field.

We simulate $N$ Lennard-Jones particles in a cubic simulation box with length $L/\sigma=10$ and periodic boundary conditions.
Here, $\sigma$ is the length scale of the Lennard-Jones potential.
We use a cutoff distance for the interparticle potential $r_c/\sigma=2.5$.
The energy parameter of the Lennard-Jones potential $\epsilon$ acts as our energy scale. 
Our time scale is $\tau=\sigma^2\gamma/\epsilon$ and 
we work at constant (supercritical) temperature $k_BT/\epsilon=1.5$.

The number of particles $N$ is chosen uniformly in the interval $200\le N \le800$.
Hence, the bulk density in the training set varies within the interval $0.2\le\rho_b\sigma^3\le0.8$.
The particles are randomly initialized in the simulation box and then equilibrated in bulk (without external force) for a total time $t/\tau=10$.
We then switch on an external force and wait $100\tau$ for the system to reach a steady state.
As shown in the Appendix of Ref.~\cite{delasHeras2023}, the initial state has no influence at all on the final steady state (note that we consider ergodic fluids only and stay away from phase transitions).
The particle initialization affects only the dynamical path followed by the system towards the steady state (provided that the system is ergodic).
Once the system is in steady state, we sample the one-body fields of interest as an average over time during at least $3\cdot10^3\tau$.

The external force field is generated randomly according to
\begin{equation}
	\fext(z)=\sum_{\alpha} f_{{\rm ext},\alpha}(z)\hat{\vecp e}_\alpha,\label{eq:fextrg}
\end{equation}
with $f_{{\rm ext},\alpha}$ being the Cartesian $\alpha$-component of the external force field and $\hat{\vecp e}_\alpha$ the unit vector along the $\alpha$-direction.
To make the model as general as possible while avoiding the sampling of multidimensional data, we use external force fields with components along the three spatial directions but allow inhomogeneity to occur only along the $z$-direction ($\hat{\vecp e}_z$).
This planar geometry allows us to work solely with one-dimensional histograms, which eases sampling, data preparation and neural network construction. 
Each component of the external force field is generated randomly via superposition of Fourier modes
\begin{equation}
	f_{{\rm ext},\alpha}(z)=a_0^\alpha+\sum_{m=1}^{M_\alpha}a_m^\alpha\sin\left(\frac{2\pi z}Lk_m^\alpha+\psi_m^\alpha\right).
	\label{eq:fextalpha}
\end{equation}
The coefficient $a_0^\alpha$, which plays a major role determining the average current, is chosen uniformly in the interval $[0,50\epsilon/\sigma]$.
The maximum number of superimposed Fourier modes for component $\alpha$ is $M_\alpha$ which is randomly selected between one and four.
The amplitudes $a_m^\alpha$ and the phases $\psi_m^\alpha$ of each Fourier mode are chosen uniformly; 
the phases within the interval $[0,2\pi)$ and the amplitudes within the interval $[0,30\epsilon/\sigma]$ for $\alpha=x,y$ and $[0,4\epsilon/\sigma]$ for $\alpha=z$.
To reduce ergodicity problems during the generation of the training set, the maximum amplitude is smaller along the inhomogeneous direction.
The parameters $k_m^\alpha$ are randomly selected integers ranging from one to four. 

\begin{figure*}
	\centering
        \includegraphics[width=1.0\linewidth]{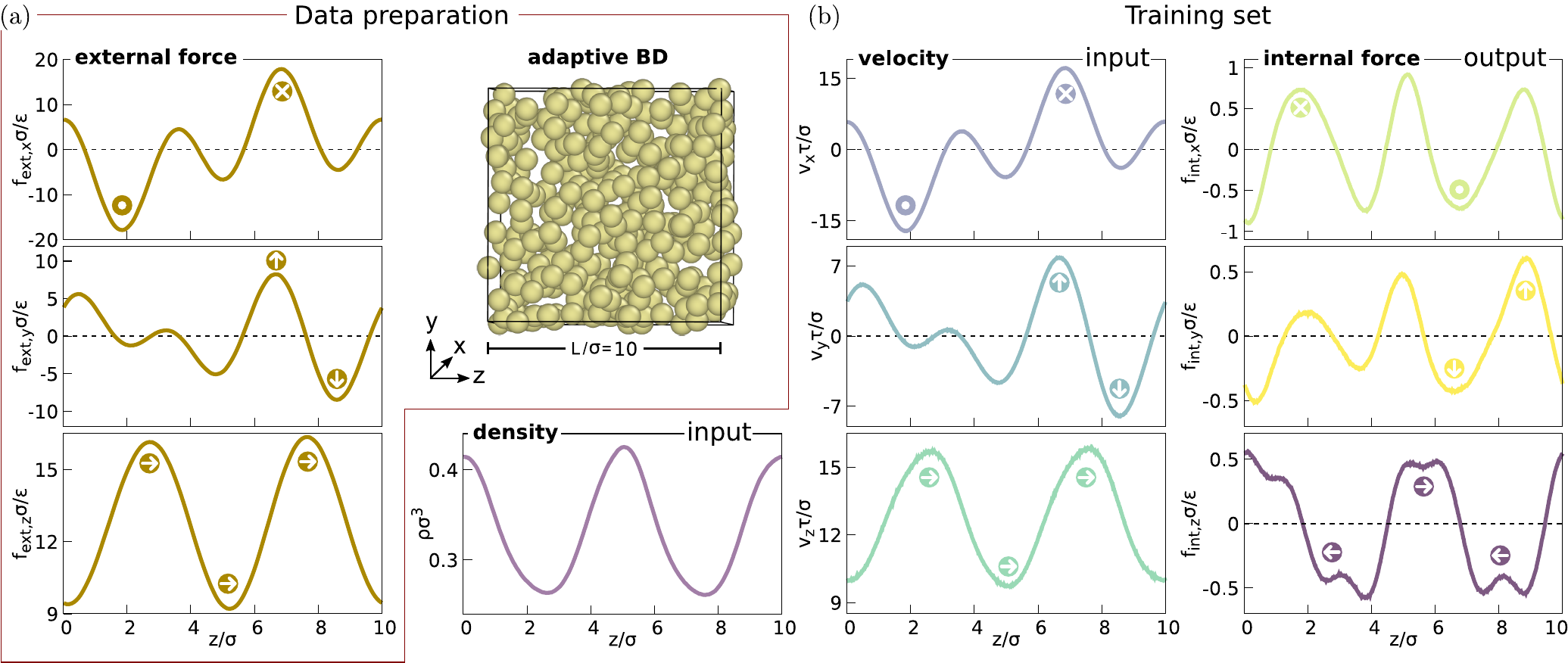}%
	\caption{{\bf Training set: from particles to fields.} 
	(a) Components of a randomly generated external force as a function of the $z$ coordinate and characteristic snapshot of the system in the corresponding steady state.
	 We generate data for the one-body fields from the many-body system by sampling the steady state which results from applying a randomly generated external force field.
	(b) One-body fields involved in the training set as sampled in the simulations.
	The fields correspond to the steady state of the external force shown in panel (a).
	Input fields: density $\rho$ and velocity $\vel$ profiles as a function of $z$.
	Output field: internal force $\fint$ as a function of $z$.
	Zero baselines in plots of vector components are indicated with horizontal dash-lines.
	The colored arrows indicate the direction at selected positions of the component of the vector field.
	}
	\label{fig1}
\end{figure*}

We then generate several external forces according to Eq.~\eqref{eq:fextalpha} and run the corresponding simulations.
Approximately five percent of the simulations were discarded since they contained regions in which the density was locally very low (smaller than $0.01\sigma^3$) which could be an indication of the system not being ergodic.
This can happen if e.g.~the external force varies strongly in a small region effectively trapping the particles. 
We complement the training set with that of Ref.~\cite{delasHeras2023}, which consists of $1000$ simulations in which the external force is generated according to Eq.~\eqref{eq:fextalpha} but  where the only non-vanishing external force contribution is the $z-$component,
i.e.~$\fext=f_{{\rm ext},z}(z)\hat{\vecp e}_z$.
Roughly ten percent of this subset of simulations corresponds to the relevant case of equilibrium systems for which $\fext$ has only a $z-$component and $a_0^z=0$.
Having equilibrium profiles in the training set is relevant to e.g.~split the internal force field into adiabatic and superadiabatic components~\cite{Fortini2014}.

We illustrate in Fig.~\ref{fig1} the data preparation and sampling of the one-body fields with one example of the training set.
The one-body fields are sampled in $z$-direction using histograms with bin size $0.01\sigma$.
The final training set contains $2897$ individual simulation results which we split as usual into training ($1877$), validation ($868$), and test ($152$) sets.
The training set is provided in the Supplementary Material.
After training, additional simulation data is obtained to compare with the predictions of the neural network for selected systems (see Sec.~\ref{Sec:results}).

\subsection{Data augmentation}\label{sec:da}

We use coordinate transformations which keep the non-equilibrium physics unchanged to augment the training data.
For instance, one can flip the $z$-axis, i.e.~perform the transformation $z\rightarrow -z$, and hence duplicate the training data by changing the original data accordingly (which in this case also implies flipping the sign of the $z$-components of the vector fields $\vel$ and $\fint$).
We use eight coordinate systems: the original one, three in which we flip only one axis, other three in which we flip two axes, and one with all axes flipped.
Moreover, since the system is inhomogeneous along the $z$-direction only, we can interchange the $x$ and $y$-components of $\vel$ and $\fint$.
Applying this symmetry to each of the eight coordinate systems results in sixteen possible coordinate transformations to augment our training data.

Data augmentation serves two purposes.
It expands the number of input-output pairs available for training (no data augmentation is used in either the validation or the test sets).
More importantly, data augmentation indirectly imposes the physically relevant symmetries of our system into the neural network.
Alternatively, it might be possible to incorporate the symmetries of the system using equivariant neural networks~\cite{Cohen2016} and other physics-informed machine learning techniques~\cite{Karniadakis2021}.

\begin{figure}
	\centering
        \includegraphics[width=0.6\linewidth]{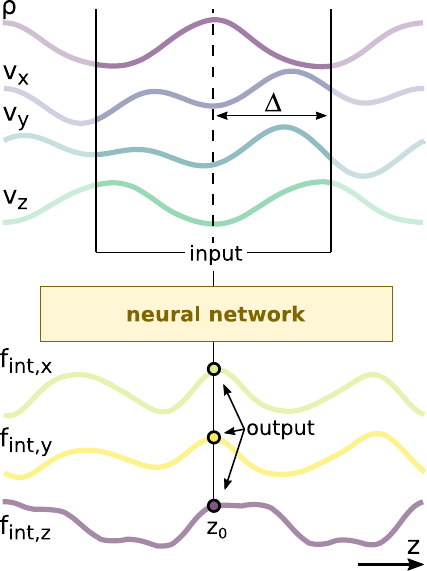}%
	\caption{{\bf Schematic of the local learning approach.}
	The input fields are the density profile $\rho$ and the three components of the velocity profile $\vel$ in an interval centered at $z_0$ and half-width $\Delta=2.8\sigma$.
        The output of the network are the three components of the internal force field at position $z_0$.
	The kinematic map is represented with a convolutional neural network.
	}
	\label{fig2}
\end{figure}

\subsection{Local learning}
The internal force field at a given position $z_0$ is fully determined by the density and the velocity profiles in the neighbourhood of that position.
This allows us to use a local learning approach, similar to the one used by Sammüller {\it et al.}~\cite{Sammuller2023} for equilibrium systems.
We use the network to represent the kinematic mapping at position $z_0$, see Fig.~\ref{fig2}.
That is, the output of the neural network is $\fintn(z_0)$.
We use here and in the following the superscript $\star$ to indicate quantities that have been obtained with the neural force functional.
The input fields of the network are $\rho$ and $\vel$ in an interval of width $2\Delta$ centered at $z_0$.
Here, we use $\Delta=2.8\sigma$ which produces the best results for our training set. 
Overfitting starts to occur for larger values.
The optimal value of $\Delta$ might depend on several variables including the interparticle potential, the temperature, and the range of currents in the training set.

Instead of one input-output training sample per simulation, the local learning strategy facilitates to use the data of each histogram bin ($10^3$ per simulation) individually, thereby increasing data efficiency substantially.
Additionally, data augmentation multiplies the number of samples by a factor sixteen, see Sec.~\ref{sec:da}.
Hence, from the original $1887$ simulations used for training we generate approximately $3\cdot10^7$ local input-output pairs.

This local learning approach is crucial since (i) it simplifies the learning process,
(ii)  it imposes the short-ranged spatial dependence of the underlying functional by construction,
and (iii) it allows us to apply the network to systems much larger than the original simulation box.
The implementation of local learning eliminates the intrinsic length scale constraint dictated by the size of the simulation box. 
Note that local learning is possible due to the specific form of the mapping that is represented.
Local learning is not possible if one learns e.g.~the relation between $\fint$ and $\fext$ since the internal force field at a given position depends on the global shape of the external field.

We use a convolutional neural network with three convolutional layers, see details in Appendix~\ref{appendix}.
The input data comprises only the required elements to generate the output, without any superfluous or lacking components.
We believe this contributes to a successful learning process and makes it possible to efficiently learn a non-trivial problem in statistical physics with a relatively simple network architecture (compared with cutting-edge models).

\section{Results}\label{Sec:results}
After training, the network acts as a neural force functional representing the kinematic functional mapping from the density and velocity profiles to the internal force field.
Combined with the force balance and the continuity equations, the neural force functional allows us to study the many-body dynamics at the one-body level.
We show several applications in this section.

\subsection{Neural custom flow}\label{sec:cf}
Custom flow is a numerical method to find the external force field that generates the desired dynamics (prescribed by the one-body density and velocity profiles).
The method has been developed for Brownian~\cite{delasHeras2019} and Newtonian~\cite{Renner2021} many-body systems.
We have used custom flow to design flows that facilitate the analysis of the superadiabatic forces~\cite{delasHeras2020,Renner2022} and to 
perform the adiabatic construction~\cite{Fortini2014} (that is, to find the conservative potential that generates in equilibrium the same density profile as that in an out-of-equilibrium system).
The adiabatic construction can be used to split the internal force into adiabatic and superadiabatic contributions, see Eq.~\eqref{eq:fadsup}. 

Custom flow, which is an example of the growing field of inverse design in statistical physics~\cite{Miskin2015,Sherman2020,Coli2022}, is computationally expensive.
The method requires to run several simulations to iteratively find the precise form of the external force field (up to the imposed numerical tolerance).
In contrast, with the network we perform the same task instantly.
We fix $\rho(\rv)$ and $\vel(\rv)$ in steady-state and use the neural network to infer the corresponding $\fintn(\fd)$.
Then, using the exact force balance equation~\eqref{eq:fb}, we solve for the external force field
\begin{equation}
	\fextn(\rv)=\gamma\vel(\rv)+k_BT\nabla\ln\rho(\rv)-\fintn(\rv).\label{eq:fext}
\end{equation}

An example of inverse design using custom flow is shown in Fig.~\ref{fig3}.
Generalizing one of the flows considered in Ref.~\cite{delasHeras2023}, we prescribe a density profile with a kink, see Fig.~\ref{fig3}(a), and also include kinks in the $x$- and $y$-components of $\vel$, see Fig.~\ref{fig3}(b).
In planar geometry, the continuity equation~\eqref{eq:continuity} imposes that in steady state the $z$-component of $\vel$ is determined up to a multiplicative constant by the density profile
\begin{equation}
v_z(z)=J_0/\rho(z),
\end{equation}
with $J_0$ being the magnitude of the steady current.
The corresponding external force is shown in Fig.~\ref{fig3}(c).
To test the validity of the method, we next run Brownian dynamics simulations using the predicted external force profile and sample the one-body fields, see the symbols (simulations) and solid-lines (neural force functional) in Fig.~\ref{fig3}.
Both sets of profiles agree well, specially considering that this is a demanding test because profiles with kinks were not included during training.
Moreover, when comparing the internal force fields obtained with BD and the neural force functional there are two main sources of error.
First, the network produces an error calculating the internal force field.
Second, due to this error in $\fintn$, the external force calculated with the force balance equation~\eqref{eq:fext} is not exactly the one that generates the prescribed kinematic profiles in simulations.
Hence, the internal force field sampled in the simulations does not exactly correspond to the internal force field of the prescribed kinematic fields.

The kink in the density profile creates a discontinuity in the ideal gas diffusive term of the force balance equation, see Eq.~\eqref{eq:diff}.
The discontinuity can only be balanced by the external force, see Eq.~\eqref{eq:fext} and Fig.~\ref{fig3}(d), because both the velocity and the internal force field cannot be discontinuous.
This is another advantage of having the internal force as the output field since its properties remain well-behaved even for demanding input fields
(besides the benefit of being able to use a local learning approach).

\subsection{Superadiabatic forces}\label{sec:split}
\begin{figure*}
	\centering
        \includegraphics[width=1.\linewidth]{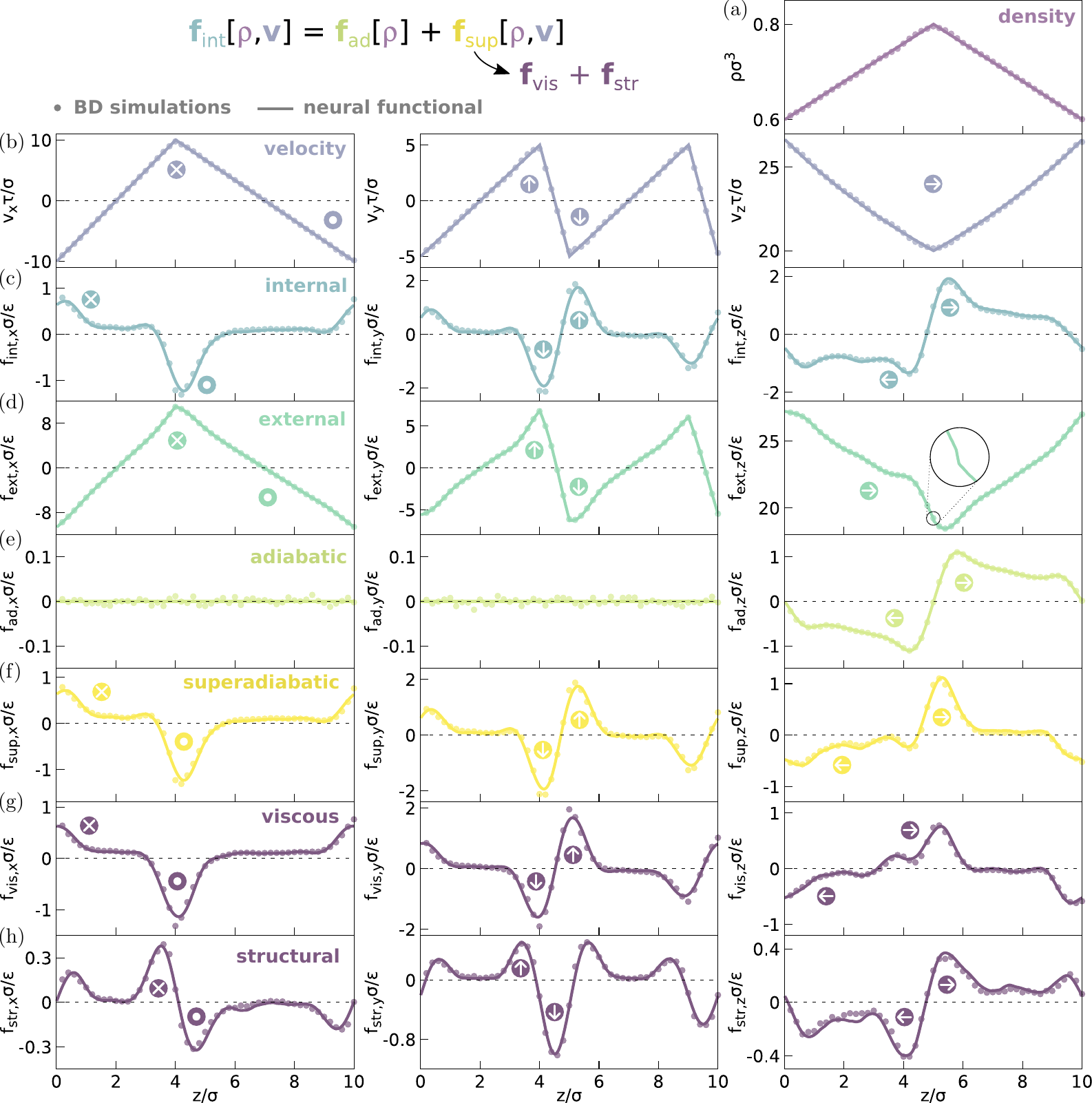}%
	\caption{
		{\bf Inverse design via neural custom flow and superadiabatic forces.}
		One-body fields according to the neural force functional (solid lines) and simulations (symbols) as a function of $z$:
		(a) density $\rho$,
		(b) velocity $\vel$,
		(c) internal force $\fint$,
		(d) external force $\fext$,
		(e) adiabatic force $\fad$,
		(f) superadiabatic force $\fsup$,
		(g) viscous superadiabatic force $\fvis$,
		(h) structural superadiabatic force $\fstr$.
		The first, second and third column correspond to the $x$-, $y$-, and $z$-components of the respective vector fields.
		The arrows indicate the direction at selected positions of the component of the vector field.
		The horizontal dashed lines indicate the zero baseline of the vector components.
		The inset in (d) is an enhanced view of the region in which $\fext$ is discontinuous.
	}
	\label{fig3}
\end{figure*}
To split the internal force field into adiabatic and superadiabatic components, we simply evaluate the neural force functional to yield the internal force field in equilibrium ($\vel=0$) which corresponds to the adiabatic force
\begin{equation}
	\fadn(\rv;[\rho])=\fintn(\rv;[\rho,\vel=0]).\label{eq:ad}
\end{equation}
The superadiabatic contribution is obtained by subtracting the adiabatic part from the total internal force field
\begin{equation}
	\fsupn(\fd)=\fintn(\fd)-\fadn(\rv;[\rho]).
\end{equation}

In simulations, we sample the adiabatic force field using the external force field obtained from Eq.~\eqref{eq:fext} with $\fadn$ as the internal force field and $\vel=0$.
The resulting force field is conservative since the adiabatic system is in equilibrium.
In Fig.~\ref{fig3}(e) and Fig.~\ref{fig3}(f) we compare the adiabatic and superadiabatic contributions obtained with the neural force functional and with simulations.
The $z$-component of the adiabatic force field is the only non-vanishing component because the equilibrium internal force field is a gradient field, see Eq.~\eqref{eq:finteq}, and our system is homogeneous along the $x-$ and $y-$directions.

\noindent{\bf Viscous and structural components.}
In steady-state, the superadiabatic force field can be split into viscous (or flow), $\fvis$, and structural, $\fstr$, components
\begin{equation}
	\fsup(\rv;[\rho,\vel])=\fvis(\rv;[\rho,\vel])+\fstr(\rv;[\rho,\vel]).\label{eq:split1}
\end{equation}
We outline here only the main ideas and refer the reader to Ref.~\cite{delasHeras2020} for a complete description.
The viscous component responds to the direction of the velocity field and it often opposes the flow.
The structural component responds to the shape of the velocity field but not to its direction and it can create structure in the fluid.
To split $\fsup$ we create a reverse steady state in which the flow points in opposite direction, $\vel_r(\rv)=-\vel(\rv)$, but the density profile remains unchanged, $\rho_r(\rv)=\rho(\rv)$.
The subscript $r$ refers to quantities in the reverse steady state.
Hence, the viscous component flips sign in the reverse system whereas the structural component remains unchanged.
The superadiabatic force field in the reverse system is then
\begin{equation}
	\fsupr(\rv;[\rho_r,\vel_r])=-\fvis(\rv;[\rho,\vel])+\fstr(\rv;[\rho,\vel]).\label{eq:split2}
\end{equation}
Note that if the superadiabatic force field is expanded in powers of the velocity field, then the odd (even) terms in powers of $\vel$ generate viscous (structural) force contributions.
From Eqs.~\eqref{eq:split1} and~\eqref{eq:split2} it follows that
\begin{eqnarray}
	\fvis(\rv;[\rho,\vel])&=&\frac{\fsup(\rv;[\rho,\vel])-\fsupr(\rv;[\rho_r,\vel_r])}2,\\
	\fstr(\rv;[\rho,\vel])&=&\frac{\fsup(\rv;[\rho,\vel])+\fsupr(\rv;[\rho_r,\vel_r])}2.
\end{eqnarray}

The superadiabatic force field in the reverse system can be obtained using the neural network according to
\begin{equation}
	\vecp f_{{\rm sup},r}^{\,\star}(\rv;[\rho_r,\vel_r])=\fintn(\rv;[\rho,-\vel])-\fadn(\rv;[\rho]).
\end{equation}

The viscous and the structural components of the superadiabatic force field (neural force functional and simulations) are shown in Fig.~\ref{fig3}(g) and Fig.~\ref{fig3}(h), respectively.
To find the external force field that in the simulations generates the reverse system, we use Eq.~\eqref{eq:fext} with $\fintn(\rv;[\rho,-\vel])$ as the internal force field.

\subsection{Transport coefficients}\label{sec:transport}

We can also use the neural force functional to effortlessly extract transport coefficients such as the shear and bulk viscosities.
We illustrate the process with two model shear and bulk flows.

\subsubsection{Shear flow}
\begin{figure*}
	\centering
        \includegraphics[width=1.0\linewidth]{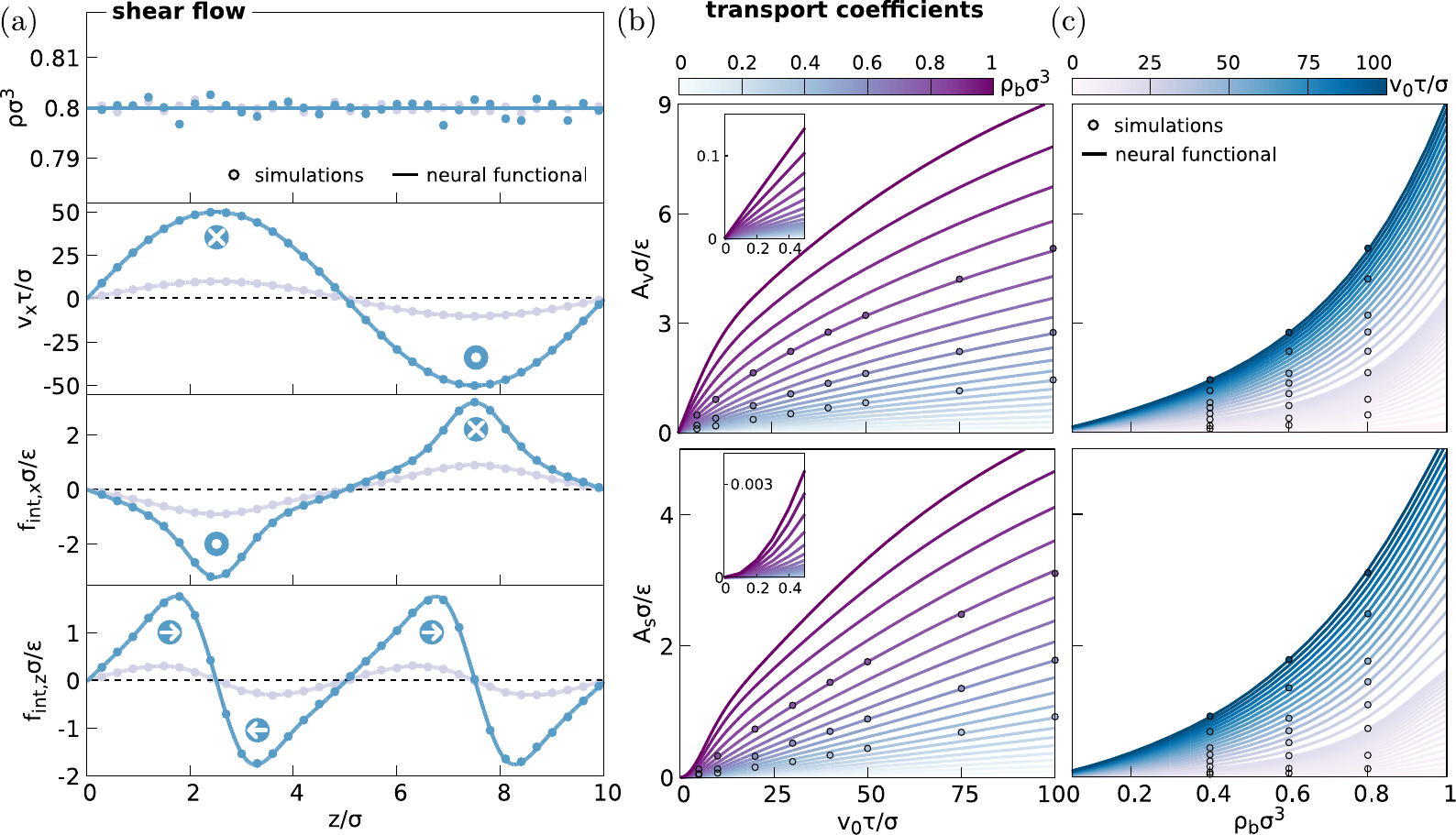}%
	\caption{
	{\bf Shear flow.}
	(a) Prescribed density $\rho$ and velocity $v_x$ profiles as well as the predicted viscous ($f_{\rm int,x}$) and structural ($f_{\rm int,z}$) components of the internal force field $\fint$ as a function of $z$ in a model shear flow.
	Solid lines (symbols) are neural force functional predictions (Brownian dynamics simulations).
	The average density is set to $\rho_b\sigma^3=0.8$ and the imposed flow is a sinusoidal wave of amplitude $v_0\sigma/\tau=50$ (blue) and $10$ (grey).
	Amplitudes of the viscous $A_v$ and the structural $A_s$ superadiabatic forces (b) as a function of the strength of the flow $v_0$ for several values of the average bulk density $\rho_b$ (color bar),
	and (c) as a function of the bulk density $\rho_b$ for several amplitudes of the flow $v_0$ (color bar).
	The amplitudes are obtained as half of the difference between the maximum and the minimum value of the corresponding force.
	The insets in (b) are a close view of the weak flow regime.
	}
	\label{fig4}
\end{figure*}

We prescribe the following spatially periodic shear flow (Kolmogorov flow~\cite{Obukhov1983})
\begin{eqnarray}
	\rho(z)&=&\rho_b,\label{eq:rhocte}\\
	\vel(z)&=&v_0\sin(2\pi z/L)\hat{\vecp e}_x,
\end{eqnarray}
with $\rho_b$ and $v_0$ constants.
That is, the density profile is uniform and the velocity profile is a sinusoidal wave along the $x$-direction.
The velocity profile is therefore divergence-free $\nabla\cdot\vel=0$ and it has a non-vanishing curl $\nabla\times\vel\neq0$.
The one-body fields are shown in Fig.~\ref{fig4}(a).

We have studied this flow in a two-dimensional system in Ref.~\cite{delasHeras2020} using custom flow to understand the superadiabatic forces.
Also, a closely related flow was analysed in Refs.~\cite{Stuhlmuller2018,Sammller2023}.
There, a sinusoidal external force drives the system.
For sufficiently weak external driving, the velocity profile closely resembles the sinusoidal shape of the driving force and the density is almost uniform.
However, for strong enough driving the velocity profile can differ substantially from a sinusoidal wave,
and the particles migrate to the regions of low shear rate generating therefore a density modulation~\cite{Stuhlmuller2018,Sammller2023}.

Here, we impose the density profile to be homogeneous, Eq.~\eqref{eq:rhocte}.
Hence, the entire internal force field is superadiabatic.
Another advantage of this flow is the natural splitting of the viscous and the structural components along different Cartesian directions~\cite{delasHeras2020}.
The internal force field along the $x$-direction (parallel to the flow) is viscous and it opposes the flow, see Fig.~\ref{fig4}(a).
This component of the force reverses if we reverse the flow.
Along the $z$-direction, perpendicular to the flow, there is additionally a superadiabatic force field which is of structural type.
Reversing the flow does not alter this component of the force.
If we remove the constraint of uniform density, the structural force field would create a density modulation~\cite{Stuhlmuller2018,Sammller2023} with peaks around the regions of low shear rate.
Note how the structural force field, $\fstr(\rv)=f_{{\rm int},z}(\rv)\hat{\vecp e}_z$, tries to move particles from the high ($z/\sigma=0$ and $5$) to the low ($z/\sigma=2.5$ and $7.5$) shear rate regions.
An external force field in $z-$direction balances the structural force such that the density profile remains homogeneous.

To simulate the flow, we use the external force field obtained with neural custom flow (as explained in Sec.~\ref{sec:cf}) and then sample the one-body fields.
The predictions of the neural force functional agree well with computer simulations, see Fig.~\ref{fig4}(a).

We have studied this flow previously in Brownian~\cite{Stuhlmuller2018,delasHeras2020,Sammller2023} and Newtonian~\cite{Renner2022} systems.
However, we used either a sinusoidal external force field~\cite{Stuhlmuller2018,Sammller2023} (in which case the velocity profile is not a perfect sinusoidal wave and 
there is a density modulation) or custom flow which is computationally demanding~\cite{delasHeras2020}.
In contrast, using the network it is straightforward to design the flow and also to obtain and analyze the internal force field.
For example, we show in Fig.~\ref{fig4}(b) and Fig.~\ref{fig4}(c) the amplitudes $A_v$ and $A_s$ of the viscous and structural superadiabatic force fields as a function of the amplitude of the flow $v_0$ and the bulk density $\rho_b$, respectively.
The amplitude is determined by taking half of the difference between the maximum and minimum values of the superadiabatic force field.
These amplitudes are (up to factors of $\rho_b$) the transport coefficients associated to the viscous and the structural response~\cite{Stuhlmuller2018}.
In the limit of weak flows $v_0\rightarrow0$ the viscous force grows linearly with $v_0$ whereas the structural force grows quadratically, see insets of Fig.~\ref{fig4}(b). 
This behaviour agrees with the superadiabatic response predicted by an approximated power functional theory constructed by expanding the excess term in powers of the velocity field~\cite{Stuhlmuller2018}.
In this geometry, the superadiabatic response generated by the first two terms in the expansion is
\begin{equation}
	\rho_b\fsup=-\eta_s\nabla\times(\nabla\times\vel)-\chi\nabla\left(\nabla\times\vel\right)^2.\label{eq:supresponse}
\end{equation}
The first term on the right hand side is the viscous component (linear in $\vel$) and the second term is the structural component (quadratic in $\vel$).
The parameters $\eta_s$ (shear viscosity) and $\chi$ are the transport coefficients associated to each superadiabatic force field in the limit of weak flows (recall that we have retained only the first terms of the expansion).
Hence, our analysis with the neural force functional confirms the suitability of previous analytical approaches.
For strong flows the network predicts the saturation of the superadiabatic forces which is also in agreement with previous simulations~\cite{nikolai}.

\subsubsection{Shear plus constant flow}
In equilibrium, the neural functional~\cite{Sammuller2023} satisfies exact sum rules that follow from Noether invariance~\cite{Hermann2021} even though they have not been imposed during training. 
In non-equilibrium, the network also complies with symmetries of the underlying physical system that have not been imposed by e.g.~data augmentation.
For example, we have verified that adding a constant to the $x$- or $y$-component of the velocity field in the shear flow shown in Fig.~\ref{fig4} leaves the internal force field predicted by the network unchanged.
This was expected because all one-body fields are homogeneous in both $x$- and $y$-directions.
Hence, we are simply changing the Galilean reference frame and the internal force field is invariant under such transformation.

\begin{figure*}
	\centering
        \includegraphics[width=1.0\linewidth]{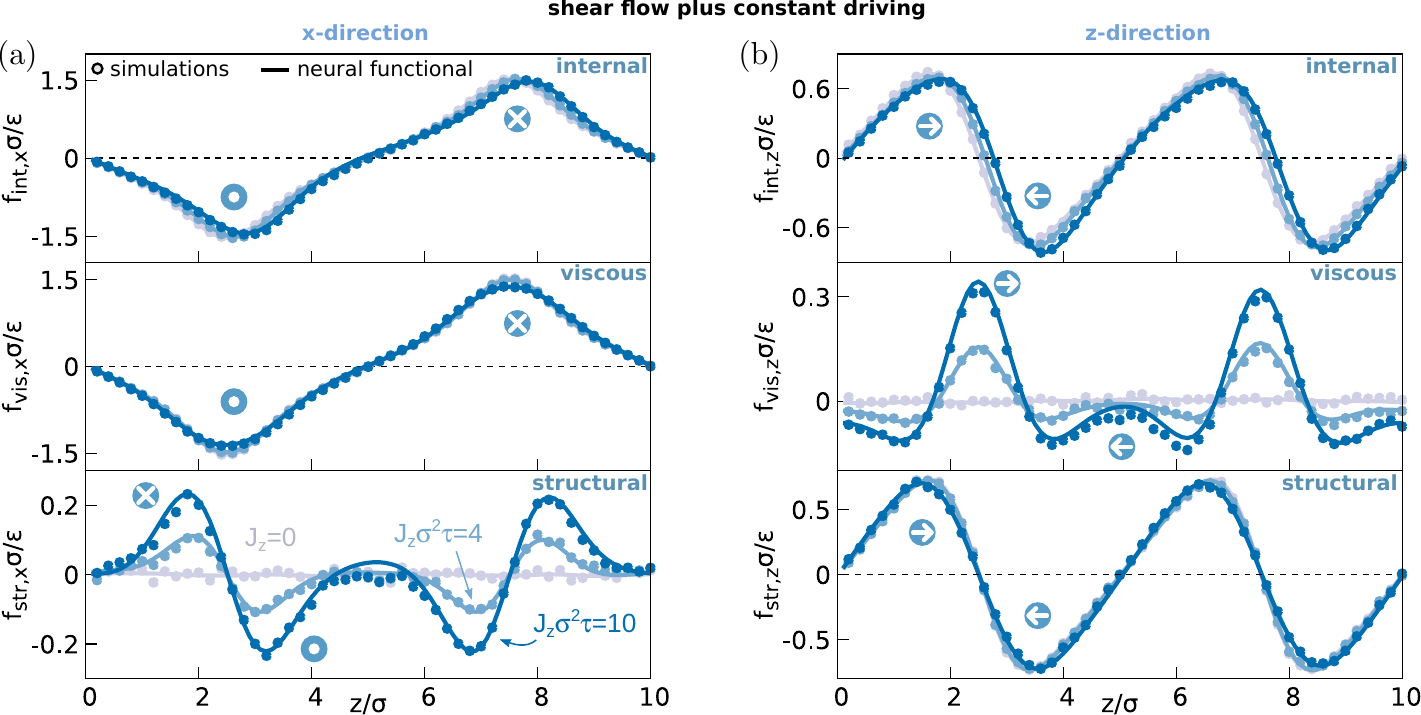}%
	\caption{{\bf Shear flow plus constant driving.} Internal forces in a system with constant density profile, $\rho_b\sigma^3=0.7$, following a sinusoidal shear flow along the $x$-direction, $v_x(z)=v_0\sin(2\pi z/L)$ with $v_0\tau/\sigma=25$, and a constant flow along the $z$-direction $v_z=v_1$ of magnitude $v_1=0$ (gray), $J_z\sigma^2\tau=4$ (light blue), and $J_z\sigma^2\tau=10$ (dark blue), with $J_z=v_1\rho_b$ the current along the $z$-direction.
	The internal force field (top row) is entirely superadiabatic $\fint=\fsup$ since $\rho$ is constant.
	The viscous $\fvis$ (middle row) and the structural $\fstr$ (bottom row) superadiabatic force fields depend on the value of $v_1$. 
	The left (right) panels display the $x$-component ($z$-component) of the forces.}
	\label{fig4b}
\end{figure*}

The velocity, the internal force field, and the external force field are inhomogeneous along the $z$-direction.
Hence, adding a constant term to the $z$-component of the velocity is not equivalent to a change between two inertial frames of reference.
The network is also able to discern whether adding a constant flow is equivalent to changing to another inertial frame of reference.
We show in Fig.~\ref{fig4b} the internal force field and its splitting into viscous and structural components for a shear flow.
Like in Fig.~\ref{fig4}, the system is constructed to have constant density and the $x$-component of the velocity profile is a sinusoidal wave.
In addition, there is a constant drift along $\hat{\vecp e}_z$.
That is
\begin{eqnarray}
	\rho(z)&=&\rho_b,\\
	\vel(z)&=&v_{0}\sin(2\pi z/L)\hat{\vecp e}_x+v_{1}\hat{\vecp e}_z.
\end{eqnarray}
Since the density is constant, the entire internal force field is superadiabatic.
The presence of the $z-$component of the velocity clearly alters the internal force field.
First, the viscous force along $\hat{\vecp e}_x$ and the structural force along $\hat{\vecp e}_z$ depend on the value of $v_1$.
Additionally, switching on the constant flow along $\hat {\vecp e}_z$ generates a superadiabatic viscous force along $\hat {\vecp e}_z$ and a structural force along $\hat{\vecp e}_x$.
Both superadiabatic force fields are not present if $v_1$ vanishes.
We have confirmed with Brownian dynamics simulations that the predictions of the neural force functional closely match the simulation data (compare lines and symbols in Fig.~\ref{fig4b}).

A steady state approximation to the excess power functional based on a series expansion in powers of the gradient of the velocity field, $\nabla\vel$, can be enough to reproduce the superadiabatic force field accurately~\cite{delasHeras2018, Stuhlmuller2018}.
This example shows that there are certain cases in which the mean value of the velocity field is also relevant to determine the superadiabatic forces.
For those cases, an approximate analytical excess power functional can contain terms that depend on the velocity field itself, $\vel$, such as those included in Ref.~\cite{delasHeras2020}.
For a discussion of exact nonequilibrium sum rules for the non-stationary dynamics, we refer the reader to Ref.~\cite{Hermann2021}.

The Galilean transformation could be also used for data augmentation to potentially improve the quality of the network predictions~\cite{Ling2016}.

\subsubsection{Bulk flow}
In the shear flow, the homogeneous density profile along with the viscous and structural forces being perpendicular to each other facilitates the analysis of the transport coefficients.
Bulk (compressible) flows are more challenging since the flow and viscous components are parallel to each other.
Nevertheless, the neural force functional still provides an accurate description of the superadiabatic forces~\cite{delasHeras2023} and the transport coefficients for bulk flows.
We illustrate this using the following flow
\begin{eqnarray}
	\rho(z)&=&\rho_0(1+\rho_1\cos(2\pi z/L)),\label{eq:rhobulk}\\
	\vecp J(z)&=&\rho(z)\vel(z)=J_0\hat{\vecp e}_z,
\end{eqnarray}
with constant $\rho_0$ (average density), $\rho_1$ (amplitude of density modulation) and $J_0$ (current).
Hence, by construction $\nabla\times\vel=0$ and $\nabla\cdot\vel\neq0$.
Since the current is constant, the velocity profile is simply $\vel(z)=J_0/\rho(z){\vecp e}_z$.

\begin{figure*}
	\centering
        \includegraphics[width=1.0\linewidth]{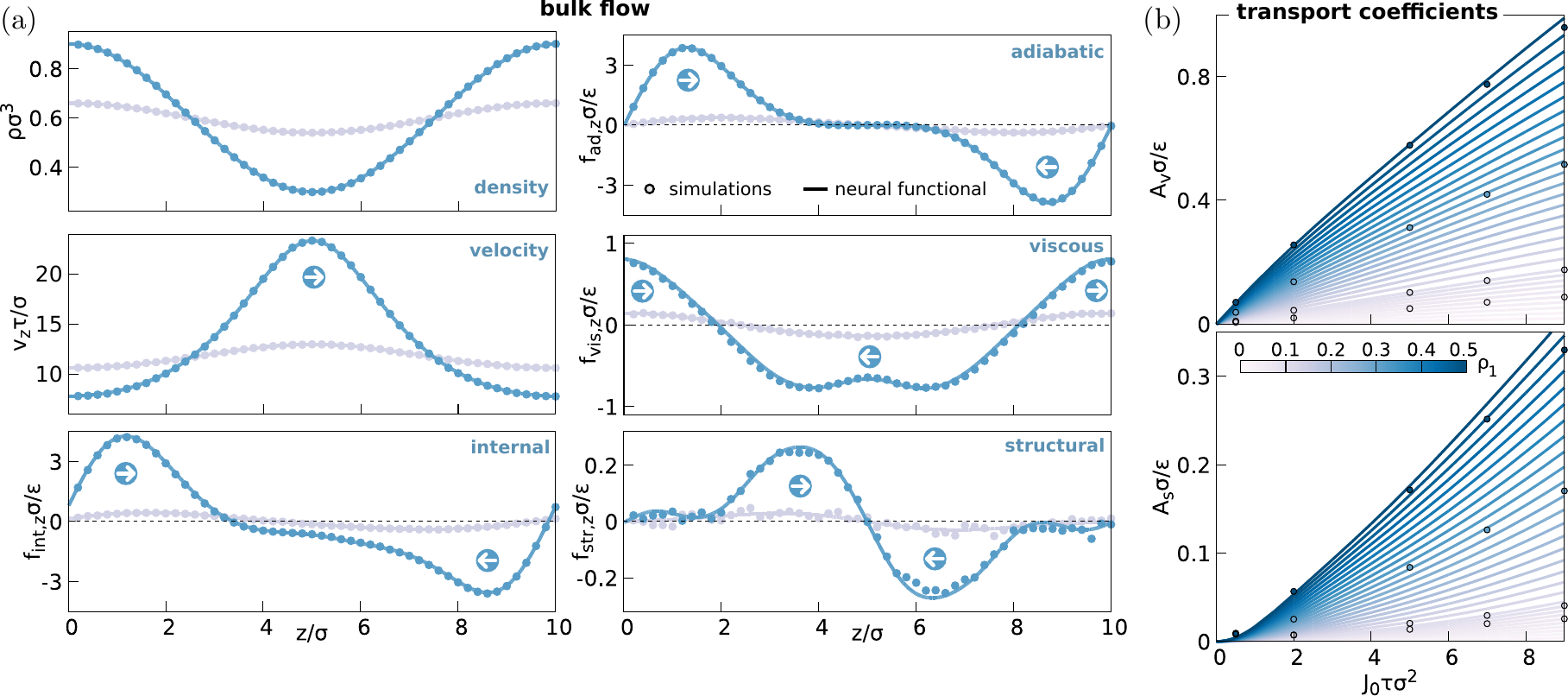}%
	\caption{{\bf Bulk flow.} (a) One-body profiles in a model bulk flow as a function of $z$ for two different density modulations: $\rho_1=0.1$ (gray) and $\rho_1=0.5$ (blue), see Eq.~\eqref{eq:rhobulk}.
	In both cases the average density is $\rho_0\sigma^3=0.6$ and the current is $J_0\sigma^2\tau=7$.
	All the vector fields point along the (gradient) $z$-direction.
	Shown are the density $\rho$, the velocity $v_z$, the internal force $f_{{\rm int},z}$, the adiabatic force $f_{{\rm ad},z}$, and the splitting of the superadiabatic force field in viscous $f_{{\rm vis},z}$ and 
	structural $f_{{\rm str},z}$ components.
	(b) Transport coefficients of the viscous ($A_v$) and the structural ($A_s$) superadiabatic forces as a function of the current $J_0$ for different values of the density modulation $\rho_1$ (color bar).
	Solid lines are neural force functional predictions and symbols are Brownian dynamics simulations.
	}
	\label{fig5}
\end{figure*}

The one-body profiles are shown in Fig.~\ref{fig5}(a) and compared to Brownian dynamics simulations.
Only the $z-$component of the internal force field does not vanish.
It contains adiabatic and superadiabatic contributions since the density is not homogeneous.
Moreover, the superadiabatic contribution incorporates both viscous and structural terms.
The complete analysis of the internal force field makes use of the network three times: for the total internal force field, for the splitting into adiabatic and superadiabatic components and a to split the superadiabatic force field into viscous and structural components.
Each time we obtain an external field  via the force balance equation and run a BD simulation to split the forces in simulations.
Although each step introduces and accumulates errors, we see in Fig.~\ref{fig5} that the network predictions remain in good agreement with the simulation results.
The neural force functional correctly describes all contributions to $\fint(\rv;[\rho,\vel])$ even though the overall shape changes non-trivially with e.g.~the amplitude of the density modulation, as illustrated in Fig.~\ref{fig5}(a).
The transport coefficients associated to the viscous ($A_v$) and the structural ($A_s$) components are shown in Fig.~\ref{fig5}(b).
They again scale differently in the limit of weak flows ($J_0\rightarrow0$) signaling a different dependence on powers of the velocity field.

\subsection{Beyond the simulation box}
The network is trained to reproduce the internal force field locally, at a given space point. 
Due to this local inference, there is no constraint on the specific choice of the system size.
The neural force functional can be used straightforwardly to predict the internal force field in systems that outscale those used during training.

\begin{figure*}
	\centering
        \includegraphics[width=0.9\linewidth]{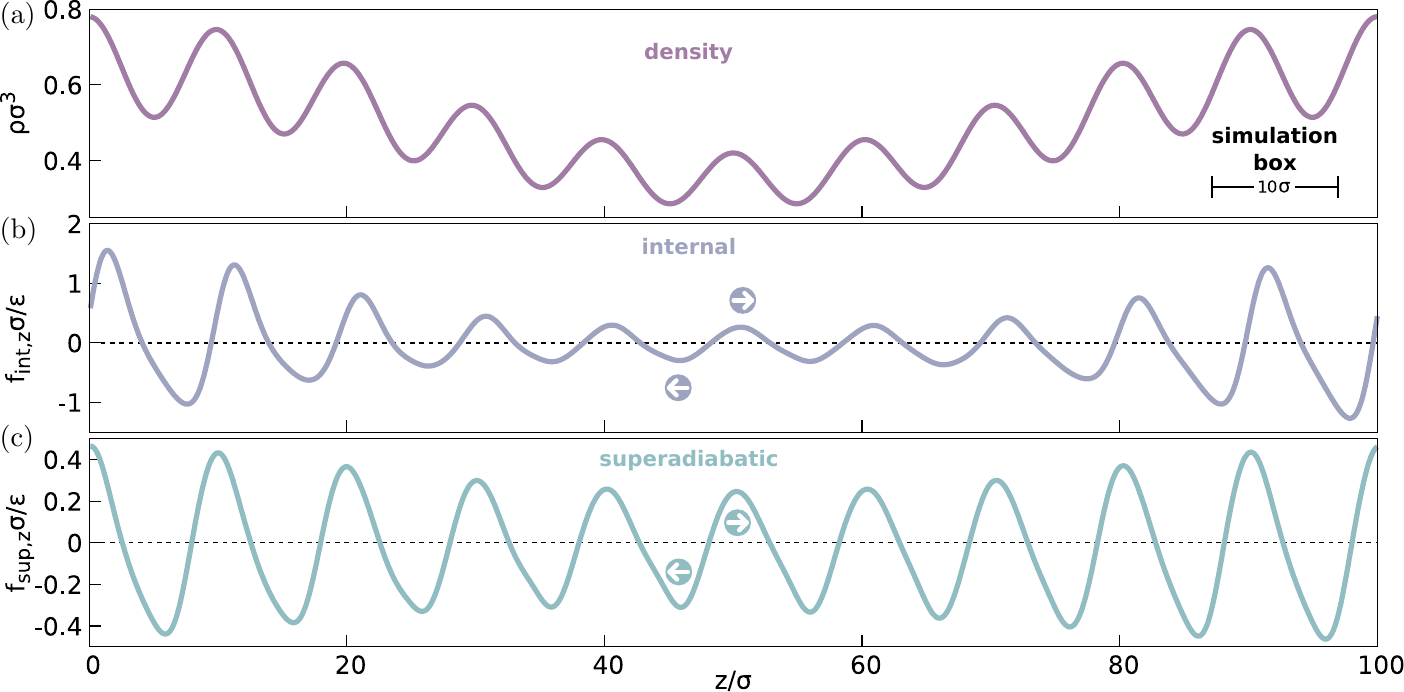}%
	\caption{
		{\bf Beyond the simulation box.}
		(a) Density profile $\rho$ and $z-$components of (b) the internal force field $\fint$ and (c) the superadiabatic force field $\fsup$ as a function of $z$ in a system of size $100\sigma$ along the $z$-direction.
		The neural network is hence used for predictions in a system which is one order of magnitude larger than those used for its training.
		The system is in steady state with a current of magnitude $J_0\sigma^2\tau=10$.
	}
	\label{fig6}
\end{figure*}

An illustrative example is shown in Fig.~\ref{fig6}.
We create a steady state in a system of size $100\sigma$ along the $z-$direction (i.e.~ten times larger than the length of the training simulation box).
The flow is directed along the $z$-axis only and the magnitude of the current is $J_0\sigma^2\tau=10$.
The density profile, and hence the velocity profile, oscillate smoothly at length scales of the order of $10\sigma$. 
Despite these rather smooth oscillations (as compared to e.g.~those in a crystal state) the superadiabatic force, Fig.~\ref{fig6}(c), is a significant contribution to the total internal force, Fig.~\ref{fig6}(b).

This system illustrates the relevance of superadiabatic forces to understand the dynamics in systems much larger than the length scale of the particles.
Our local learning approach opens the door to describe the dynamics of macroscopic systems with microscopic resolution.

\subsection{Beyond steady states}

Although the network has been trained only with steady states, we can use it to investigate full non-equilibrium situations in which the one-body fields depend explicitly on time $t$.
This constitutes an approximation to the real dynamics since the memory effects in full non-equilibrium and in steady state are not identical~\cite{Treffenstdt2020}.
Formally, the full time-dependent nonequilibrium mapping as given by PFT requires to include not only instantaneous profiles, but also their history up to the time of interest.
Nevertheless, as we demonstrate here, the agreement with simulations is still good since the network provides a reasonable approximation for the superadiabatic force field in full non-equilibrium.

In the general case, a time and space-dependent external force field $\fext(\rv,t)$ drives the dynamics of the density $\rho(\rv,t)$ and the velocity $\vel(\rv,t)$ profiles.

We discretize time in steps of $\Delta t$.
Using the continuity equation~\eqref{eq:continuity}, we evolve the density profile one time step
\begin{equation}
	\rho(\rv,t+\Delta t) = \rho(\rv,t)-\Delta t\nabla\cdot \vecp{J}(\rv,t).\label{eq:rhotdeltat}
\end{equation}
Via the exact force balance equation~\eqref{eq:fb}, we can calculate the velocity profile at time $t+\Delta t$
\begin{eqnarray}
	\gamma\vel(\rv,t+\Delta t)&=&\fint(\rv,t+\Delta t)+\fext(\rv,t+\Delta t)\nonumber\\
	&&-k_BT\nabla\ln\rho(\rv,t+\Delta t).\label{eq:iteration}
\end{eqnarray}
Since the internal force at time $t+\Delta t$ is unknown, the solution to this equation can be found using a Picard iteration:
We start with a guess for $\vel(\rv,t+\Delta t)$ and use it together with $\rho(\rv,t+\Delta t)$ from Eq.~\eqref{eq:rhotdeltat} and the network to obtain the corresponding internal force field $\fint(\rv,t+\Delta t)$.
Next, using the right hand side of Eq.~\eqref{eq:iteration} we construct a new $\vel(\rv,t+\Delta t)$ and repeat the procedure until the left and the right hand side of Eq.~\eqref{eq:iteration} coincide up to a given tolerance.
In our experience, only a few iterations are needed to achieve convergence.

In practice, if the time step is small enough (here we use $\Delta t/\tau=5\cdot10^{-5}$) the Picard iteration to evolve the velocity profile in time is not necessary. 
The velocity profile at the next time step is well approximated by assuming
\begin{eqnarray}
	\gamma\vel(\rv,t+\Delta t)&=&\fint(\rv,t)+\fext(\rv,t+\Delta t)\nonumber\\
	&&-k_BT\nabla\ln\rho(\rv,t+\Delta t),\label{eq:iteration2}
\end{eqnarray}
with
\begin{equation}
	\fint(\rv,t)=\fintn(\rv;[\rho(\rv,t),\vel(\rv,t)]),
\end{equation}
that is, the network prediction for the internal force field of the previous time step.

\noindent{\bf Neural dynamical density functional.}
To draw a comparison we construct here a neural dynamical density functional (nDDFT) which neglects the superadiabatic contributions to the internal force field.
The task is straightforward since we simply need to use $\fint(\rv,t)=\fadn(\rv,t;[\rho])$ in Eq.~\eqref{eq:iteration2}.
Recall that the adiabatic force field $\fadn(\rv,t;[\rho])$ is at each time $t$ that of an equilibrium system ($\vel=0$) with the same $\rho$ as the out-of-equilibrium system, see Eq.~\eqref{eq:ad}.

We compare predictions of the neural force functional and nDDFT with computer simulations in Fig.~\ref{fig7} and in Supplementary Movie 1.
The time-dependent one-body fields in the simulations have been obtained by averaging at the desired time $t$ over an ensemble of $\sim\!10^5$ simulations that differ in the initial microstate and in the realization of the noise (Brownian motion).
The starting point is a bulk system with constant density $\rho_b\sigma^3=0.5$. At $t=0$ we switch on the external force field shown in Fig.~\ref{fig7}(a) which is
\begin{equation}
	\fext(z)=f_0\sin\left(k_0 z\right)\hat{\vecp e}_x+f_1\cos(\omega_1 t-k_1z)\hat{\vecp e}_z,
\end{equation}
with parameters $f_0\sigma/\epsilon=30$, $k_0L=2\pi$, $f_1\sigma/\epsilon=5$, $\omega_1\tau=4\pi$, and $k_1L=4\pi$.
That is, the motion is driven by a static wave in $x$-direction, and a travelling wave in $z$-direction.

The system responds to the external driving with a density modulation, Fig.~\ref{fig7}(b), that travels along the $z$-direction.
Dynamical density functional theory is entirely unaware of the internal force field along the $x$-direction, Fig.~\ref{fig7}(c), which is non-conservative and purely superadiabatic since the density is homogeneous in this direction.
Note that the flow in $x-$direction is given according to the force balance equation~\eqref{eq:fb} by the sum of the external and the internal force fields. 
Therefore, the absence of superadiabatic forces in nDDFT leads in general to an inaccurate description of such nonequilibrium flows~\cite{delasHeras2023}, which we have exemplified here via the time-dependent situation depicted in Fig.~\ref{fig7}.
In contrast, the internal force field provided by the neural force functional and therefore the flow are in good agreement with simulations, see Fig.~\ref{fig7}(c).

The internal force field along the $z$-direction, Fig.~\ref{fig7}(d), has both adiabatic and superadiabatic components.
The adiabatic component dominates and hence the prediction of nDDFT seems at first glance reasonable as compared to simulations.
However, the superadiabatic component, which is correctly reproduced by the neural force functional, is far from being negligible and it is responsible for a clear physical effect.
We show in Fig.~\ref{fig7}(d) the amplitude of the density modulation $\Delta\rho$ as a function of time (see also Supplementary Movie 1).
The density modulation grows after switching on the external force and afterwards it varies periodically with time.
Dynamical density functional theory completely misses this effect which is due to structural superadiabatic forces generated by the flow in $z$-direction.
Again the neural force functional reproduces the oscillations of the density modulation although it slightly overestimates the amplitude.
The differences with simulation results arise, at least partially, due to memory effects being different in steady state and in full non-equilibrium, which is not captured by our neural network due to the steady state training data.

\begin{figure}
	\centering
        \includegraphics[width=0.95\linewidth]{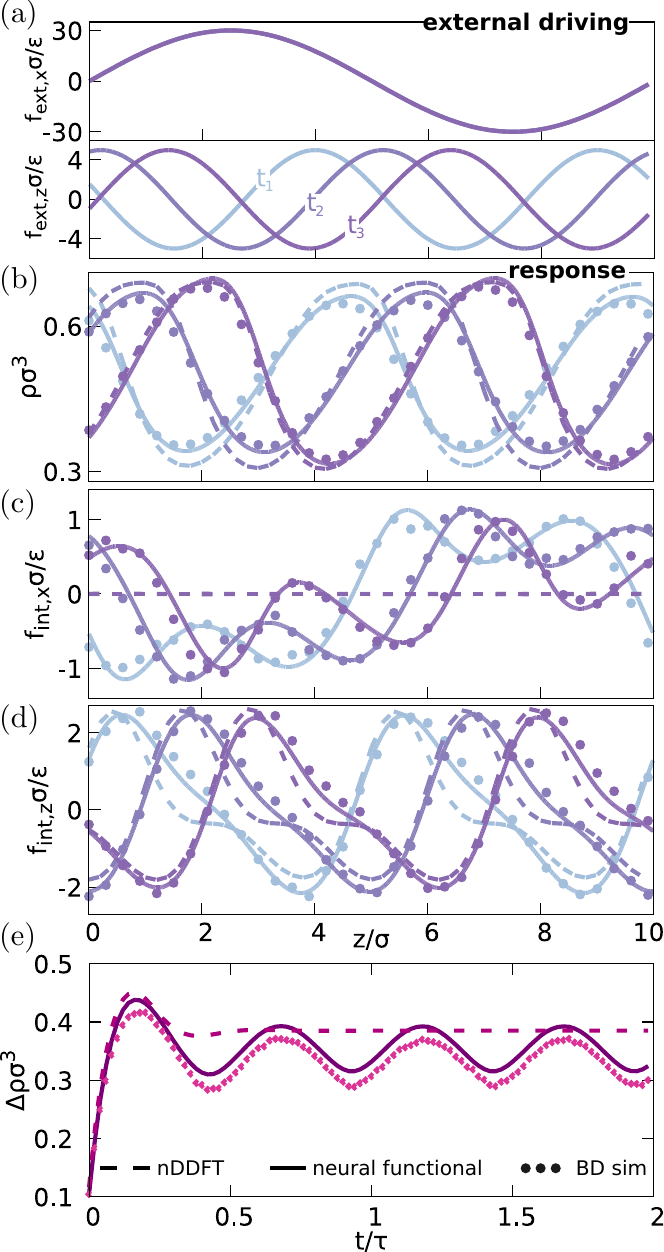}%
	\caption{{\bf Beyond steady state.} Time evolution of selected one-body fields in a full non-equilibrium system according to the neural force functional (solid lines), nDDFT (dashed lines), and BD simulations (symbols).
	The color indicates the time $t_1/\tau=0.40$ (blue), $t_2/\tau=0.64$ (soft violet), and $t_3/\tau=0.72$ (dark violet).
	Shown are as a function of the $z$ coordinate: the components of the external force field $\fext$ in $x$- and $z$-direction (a), the density profile $\rho$ (b), and the components of the internal force field in $x$- (c) and $z$-direction (d).
	Panel (e) shows the amplitude of the density modulation $\Delta\rho$ as a function of time $t$.
		}
	\label{fig7}
\end{figure}

Our neural force functional can simultaneously generalize to systems that are larger than the simulation box and that are in full non-equilibrium.
An illustrative example is shown in Supplementary Movie 2 where a system with size $L/\sigma=30$ and subject to a complex time-dependent external force is analyzed.
The predictions of the neural force functional for the density and the velocity profiles as well as for the internal force field are much closer to the simulation data than those of nDDFT.

\section{Conclusions and future work}

We have trained a neural network to represent the kinematic functional mapping $\{\rho,\vel\}\rightarrow\fint$ described in PFT for particles following overdamped Brownian dynamics.
We create the one-body fields directly from particle-based computer simulations of the supercritical Lennard-Jones fluid.
After machine learning the mapping, the network can be deployed in several applications including custom flow, the analysis of superadiabatic forces, and the quantification of transport coefficients.  
Analogously to neural functional theory in equilibrium~\cite{Sammuller2023}, the application of the network is not restricted to the original size of the simulation box in which the training data has been created.
The representation of the functional kinematic mapping is constructed in a spatially local manner and hence the prediction of the internal force field can be applied straightforwardly to systems of virtually any size.
Despite the network being trained with steady-state data, in our tests it gives a reasonable approximation of the internal force field in full non-equilibrium situations.
In our comparison with simulations, only relatively small deviations occur due to memory effects.
Note, however, that the time scales of memory effects can be small~\cite{Treffenstdt2021} and hence large deviations might appear at sufficiently small time scales.
The results agree well with simulations and clearly outperform dynamical density functional theory that can also be straightforwardly implemented as a limiting case of the neural force functional.

Several applications and extensions of this work merit further consideration. 
(i) We have trained the network at constant (supercritical) temperature $T$.
A natural extension is to prepare a training set with different values of the temperature and using $T$ as another input network parameter.
Complications might arise in e.g.~the two-phase regions of the phase diagram.
(ii) The method can be directly applied to other interparticle potentials, including hard interactions~\cite{Mederos2014} for which the internal force field can be sampled indirectly via the force balance equation.
Of particular interest are the cases of active particles~\cite{Hermann2019} and anisotropic particles.
There, the one-body density distribution depends not only on the spatial coordinates but also on the particle orientations.
Also, if there are torques acting on the particles, the force balance equation couples to a torque balance equation~\cite{Wensink2007,Renner2023} and hence, the internal torque field also needs to be learnt.
(iii) Finite size effects (and those due to truncation of the potential) might be systematically analyzed with a training set containing data from several simulation box sizes (cutoff distances).
In particular it might be possible to rationalize the effect of the lateral size of the simulation box~\cite{Chacn2006,Duque2008,Ogawa2019} on the dynamics, which constitutes an open problem that so far has not received much attention.
(iv) The network used here is simple in terms of architecture and number of trainable parameters.
Therefore, it might be feasible to generalize the geometry further and to train networks with generic two- and three-dimensional flows.
Reduced-variance sampling methods~\cite{Borgis2013,Heras2018a,Schultz2016,Rotenberg2020,Renner2023} can improve the sampling efficiency during the generation of the training set.
(v) We have used only soft external forces but the training set can be complemented with other profiles such as e.g.~particles in the presence of hard walls.
(vi) As in equilibrium~\cite{Sammuller2023}, the neural force functional can be used to perform functional calculus.
Using automatic differentiation~\cite{Margossian2019}, it is possible to compute the functional derivatives of the output with respect to the inputs of the neural network. 
This gives direct access to e.g.~the second and higher order derivatives of the excess power functional with respect to the velocity.
In the limit of weak flows, the second derivative is related to the transport coefficients such as shear and bulk viscosities.
Higher order derivatives might provide valuable information about the mathematical structure of the functional.
Also, using functional line integration~\cite{Brader2015,Sammuller2023} might be a practical route to the determination of the value of the excess power functional for prescribed $\rho(\rv)$ and $\vel(\rv)$.
(vii) Arguably the most promising extension is the application of the method to full non-equilibrium systems for which both, the training set (which then carries an explicit time dependence) and the architecture of the network need to be modified.
A neural force functional trained with full non-equilibrium data in combination with automatic differentiation might provide direct access to the memory kernels~\cite{Lesnicki2016,Jung2017,Daldrop2017,Meyer2020} of the many-body system.
(viii) Finally, the method is not restricted to overdamped Brownian dynamics.
The network can be trained with data from particles following e.g.~Newtonian~\cite{Schmidt2018}, Langevin, and quantum many-body dynamics~\cite{Schmidt2015} in full non-equilibrium.
There, the internal force field and the kinetic stress tensor carry in general an explicit dependence on the acceleration field~\cite{Brtting2019,Renner2022}. 
Hence, it might be convenient to also include the acceleration field, $\vecp a$, as an input field of the training set even though it follows from the time derivative of the velocity field ($\vecp a=\dot\vel$).
Moreover, the transport term in the force balance equation is not diffusive and also needs to be learnt.

We have trained the network with simulation data but it might be possible to generate a training set directly from experiments.
Arrays of optical tweezers~\cite{Schffner2020}, magnetic patterns~\cite{Stuhlmller2023} and micro-fabricated obstacles~\cite{Morin2016} are examples of experimental setups in which the external force field can be customized. 
The internal force field could then be measured either directly~\cite{Dong2022} or indirectly via the force balance equation and also via machine learning the non-equilibrium dynamics from time-lapse microscopy images~\cite{Gnesotto2020}.

\appendix
\section{Architecture of the neural network}\label{appendix}

\begin{figure*}
	\centering
        \includegraphics[width=0.9\linewidth]{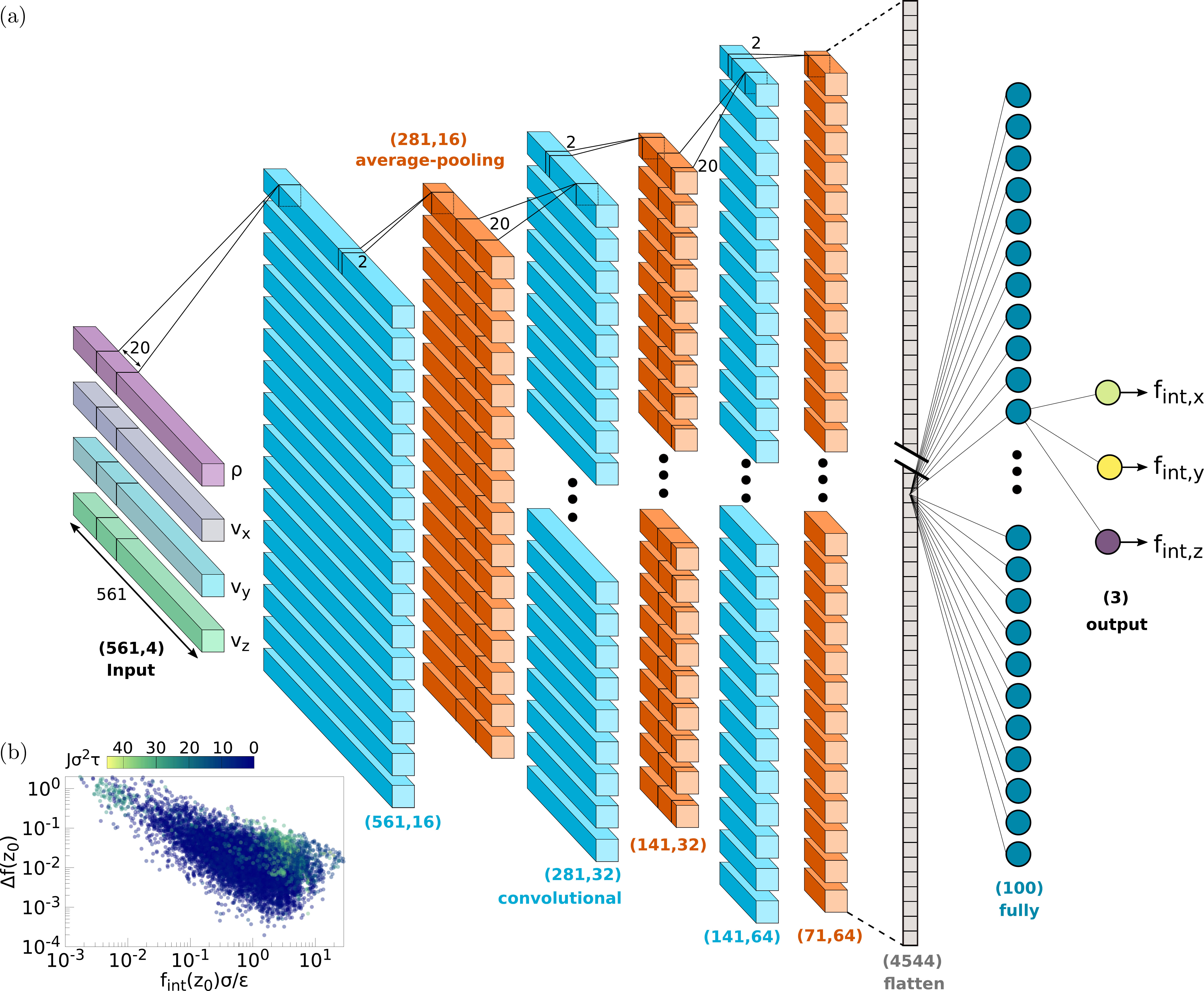}%
	\caption{{\bf Neural network.}
	(a) Schematic of the convolutional neural network.
	The input layer contains windows of the density and the three components of the velocity profile.
	Convolutional (average-pooling) layers are represented in blue (orange). The last average-pooling layer is flattened (grey)
	and connected to a fully connected layer (blue circles). The output layer consists of three nodes that output the components of $\fint$ at position $z_0$.
	The numbers inside parentheses indicate the elements per channel and the number of channels of each layer.
	(b) Log-log plot of the relative error $\Delta f(z_0)$ of the neural network as a function of the magnitude of the internal force $f_{\rm int}(z_0)$ for the samples in the test set.
	The data points are colored according to the magnitude of the current (color bar).
	}
	\label{figA}
\end{figure*}

We use a convolutional neural network, see Fig.~\ref{figA}(a) implemented in Keras~\cite{Chollet2022}.
The size of the input layer is $(561,4)$.
That is, we use $4$ channels (one-dimensional arrays) to represent the four one-body profiles (density and three components of the velocity) which are discretized within $[z_0-\Delta,z_0+\Delta]$ leading to $561$ input values per profile.
The input layer is then processed by convolutional layers followed by a fully connected layer.

The first convolutional layer uses $16$ filters.
Hence, it takes the input layer and outputs a feature map of size $(561,16)$.
Each value of the feature map is calculated by convolving the values of all the channels in the previous layer in a window of size $20$ (kernel size).

We use three convolutional layers connected sequentially and double the number of filters in each layer (the kernel size is kept constant).
An average pooling layer is connected after each convolutional layer.
The pooling layers calculate the average of two consecutive values of the feature map in the preceding convolutional layer, halving therefore the size of the feature map.

After the last pooling layer, the two-dimensional data is reshaped to a 1d-array via a flatten layer which is then connected to a fully connected layer containing $100$ nodes.
This layer connects to the output layer, made of three single nodes that output the three components of the internal force field at position $z_0$, i.e.~$\fintn(z_0)$.

We use softplus activation functions for all nodes in the network since, in our experience, they significantly improve the quality of the output.

The network contains about $5\cdot10^4$ parameters that are adjusted with the Adam optimizer to minimize the mean-square-error between network predictions and actual values of the internal force field.
The initial learning rate is $2.5\cdot10^{-4}$ and it decreases by $1.5\%$ after each epoch. 
The network is trained for $150$ epochs with an initial batch size of $512$ which is doubled every $50$ epochs.
The sequence of input-output pairs in each batch is selected randomly.

We show in Fig.~\ref{figA}(b) the relative error of the network prediction $\Delta f=|\fintn(z_0)-\fint(z_0)|/|\fint(z_0)|$ as a function of the magnitude of the internal force field $|\fint(z_0)|$ for the data in the test set.
Recall that quantities without a star are simulation data.
Data points are colored according to the magnitude of the current.
Samples with large relative errors typically correspond to cases where the internal force field is small.
A large relative $\Delta f$ is then expected since the statistical noise is comparable to the signal.
This likely explains also the general trend of negative slope in Fig.~\ref{figA}(b).
For fixed value of the magnitude of the internal force, the error increases with the magnitude of the current.
For a comparable size of the training set, higher precision is achieved in equilibrium using a neural network to represent the functional map from $\rho$ to $c_1$~\cite{Sammuller2023}.
This is not surprising due to the increased mathematical complexity of the non-equilibrium mapping. 
If higher precision is required, it should be possible to decrease the relative error by increasing the size of the training set.


\begin{thebibliography}{81}%
\makeatletter
\providecommand \@ifxundefined [1]{%
 \@ifx{#1\undefined}
}%
\providecommand \@ifnum [1]{%
 \ifnum #1\expandafter \@firstoftwo
 \else \expandafter \@secondoftwo
 \fi
}%
\providecommand \@ifx [1]{%
 \ifx #1\expandafter \@firstoftwo
 \else \expandafter \@secondoftwo
 \fi
}%
\providecommand \natexlab [1]{#1}%
\providecommand \enquote  [1]{``#1''}%
\providecommand \bibnamefont  [1]{#1}%
\providecommand \bibfnamefont [1]{#1}%
\providecommand \citenamefont [1]{#1}%
\providecommand \href@noop [0]{\@secondoftwo}%
\providecommand \href [0]{\begingroup \@sanitize@url \@href}%
\providecommand \@href[1]{\@@startlink{#1}\@@href}%
\providecommand \@@href[1]{\endgroup#1\@@endlink}%
\providecommand \@sanitize@url [0]{\catcode `\\12\catcode `\$12\catcode
  `\&12\catcode `\#12\catcode `\^12\catcode `\_12\catcode `\%12\relax}%
\providecommand \@@startlink[1]{}%
\providecommand \@@endlink[0]{}%
\providecommand \url  [0]{\begingroup\@sanitize@url \@url }%
\providecommand \@url [1]{\endgroup\@href {#1}{\urlprefix }}%
\providecommand \urlprefix  [0]{URL }%
\providecommand \Eprint [0]{\href }%
\providecommand \doibase [0]{https://doi.org/}%
\providecommand \selectlanguage [0]{\@gobble}%
\providecommand \bibinfo  [0]{\@secondoftwo}%
\providecommand \bibfield  [0]{\@secondoftwo}%
\providecommand \translation [1]{[#1]}%
\providecommand \BibitemOpen [0]{}%
\providecommand \bibitemStop [0]{}%
\providecommand \bibitemNoStop [0]{.\EOS\space}%
\providecommand \EOS [0]{\spacefactor3000\relax}%
\providecommand \BibitemShut  [1]{\csname bibitem#1\endcsname}%
\let\auto@bib@innerbib\@empty
\bibitem [{\citenamefont {L\"{o}wen}(2001)}]{Lwen2001}%
  \BibitemOpen
  \bibfield  {author} {\bibinfo {author} {\bibfnamefont {H.}~\bibnamefont
  {L\"{o}wen}},\ }\bibfield  {title} {\bibinfo {title} {Colloidal soft matter
  under external control},\ }\href
  {https://doi.org/10.1088/0953-8984/13/24/201} {\bibfield  {journal} {\bibinfo
   {journal} {J. Phys.: Condens. Matter}\ }\textbf {\bibinfo {volume} {13}},\
  \bibinfo {pages} {R415–R432} (\bibinfo {year} {2001})}\BibitemShut
  {NoStop}%
\bibitem [{\citenamefont {Erbe}\ \emph {et~al.}(2008)\citenamefont {Erbe},
  \citenamefont {Zientara}, \citenamefont {Baraban}, \citenamefont {Kreidler},\
  and\ \citenamefont {Leiderer}}]{Erbe2008}%
  \BibitemOpen
  \bibfield  {author} {\bibinfo {author} {\bibfnamefont {A.}~\bibnamefont
  {Erbe}}, \bibinfo {author} {\bibfnamefont {M.}~\bibnamefont {Zientara}},
  \bibinfo {author} {\bibfnamefont {L.}~\bibnamefont {Baraban}}, \bibinfo
  {author} {\bibfnamefont {C.}~\bibnamefont {Kreidler}},\ and\ \bibinfo
  {author} {\bibfnamefont {P.}~\bibnamefont {Leiderer}},\ }\bibfield  {title}
  {\bibinfo {title} {Various driving mechanisms for generating motion of
  colloidal particles},\ }\href
  {https://doi.org/10.1088/0953-8984/20/40/404215} {\bibfield  {journal}
  {\bibinfo  {journal} {J. Phys.: Condens. Matter}\ }\textbf {\bibinfo {volume}
  {20}},\ \bibinfo {pages} {404215} (\bibinfo {year} {2008})}\BibitemShut
  {NoStop}%
\bibitem [{\citenamefont {Menzel}(2015)}]{Menzel2015}%
  \BibitemOpen
  \bibfield  {author} {\bibinfo {author} {\bibfnamefont {A.~M.}\ \bibnamefont
  {Menzel}},\ }\bibfield  {title} {\bibinfo {title} {Tuned, driven, and active
  soft matter},\ }\href {https://doi.org/10.1016/j.physrep.2014.10.001}
  {\bibfield  {journal} {\bibinfo  {journal} {Phys. Rep.}\ }\textbf {\bibinfo
  {volume} {554}},\ \bibinfo {pages} {1–45} (\bibinfo {year}
  {2015})}\BibitemShut {NoStop}%
\bibitem [{\citenamefont {Velev}\ and\ \citenamefont
  {Bhatt}(2006)}]{Velev2006}%
  \BibitemOpen
  \bibfield  {author} {\bibinfo {author} {\bibfnamefont {O.~D.}\ \bibnamefont
  {Velev}}\ and\ \bibinfo {author} {\bibfnamefont {K.~H.}\ \bibnamefont
  {Bhatt}},\ }\bibfield  {title} {\bibinfo {title} {On-chip micromanipulation
  and assembly of colloidal particles by electric fields},\ }\href
  {https://doi.org/10.1039/b605052b} {\bibfield  {journal} {\bibinfo  {journal}
  {Soft Matter}\ }\textbf {\bibinfo {volume} {2}},\ \bibinfo {pages} {738}
  (\bibinfo {year} {2006})}\BibitemShut {NoStop}%
\bibitem [{\citenamefont {Vissers}\ \emph {et~al.}(2011)\citenamefont
  {Vissers}, \citenamefont {Wysocki}, \citenamefont {Rex}, \citenamefont
  {L\"{o}wen}, \citenamefont {Royall}, \citenamefont {Imhof},\ and\
  \citenamefont {van Blaaderen}}]{Vissers2011}%
  \BibitemOpen
  \bibfield  {author} {\bibinfo {author} {\bibfnamefont {T.}~\bibnamefont
  {Vissers}}, \bibinfo {author} {\bibfnamefont {A.}~\bibnamefont {Wysocki}},
  \bibinfo {author} {\bibfnamefont {M.}~\bibnamefont {Rex}}, \bibinfo {author}
  {\bibfnamefont {H.}~\bibnamefont {L\"{o}wen}}, \bibinfo {author}
  {\bibfnamefont {C.~P.}\ \bibnamefont {Royall}}, \bibinfo {author}
  {\bibfnamefont {A.}~\bibnamefont {Imhof}},\ and\ \bibinfo {author}
  {\bibfnamefont {A.}~\bibnamefont {van Blaaderen}},\ }\bibfield  {title}
  {\bibinfo {title} {Lane formation in driven mixtures of oppositely charged
  colloids},\ }\href {https://doi.org/10.1039/c0sm01343a} {\bibfield  {journal}
  {\bibinfo  {journal} {Soft Matter}\ }\textbf {\bibinfo {volume} {7}},\
  \bibinfo {pages} {2352} (\bibinfo {year} {2011})}\BibitemShut {NoStop}%
\bibitem [{\citenamefont {Tierno}\ \emph {et~al.}(2007)\citenamefont {Tierno},
  \citenamefont {Muruganathan},\ and\ \citenamefont {Fischer}}]{Tierno2007}%
  \BibitemOpen
  \bibfield  {author} {\bibinfo {author} {\bibfnamefont {P.}~\bibnamefont
  {Tierno}}, \bibinfo {author} {\bibfnamefont {R.}~\bibnamefont
  {Muruganathan}},\ and\ \bibinfo {author} {\bibfnamefont {T.~M.}\ \bibnamefont
  {Fischer}},\ }\bibfield  {title} {\bibinfo {title} {Viscoelasticity of
  dynamically self-assembled paramagnetic colloidal clusters},\ }\href
  {https://doi.org/10.1103/PhysRevLett.98.028301} {\bibfield  {journal}
  {\bibinfo  {journal} {Phys. Rev. Lett.}\ }\textbf {\bibinfo {volume} {98}},\
  \bibinfo {pages} {028301} (\bibinfo {year} {2007})}\BibitemShut {NoStop}%
\bibitem [{\citenamefont {Lips}\ \emph {et~al.}(2021)\citenamefont {Lips},
  \citenamefont {Stoop}, \citenamefont {Maass},\ and\ \citenamefont
  {Tierno}}]{Lips2021}%
  \BibitemOpen
  \bibfield  {author} {\bibinfo {author} {\bibfnamefont {D.}~\bibnamefont
  {Lips}}, \bibinfo {author} {\bibfnamefont {R.~L.}\ \bibnamefont {Stoop}},
  \bibinfo {author} {\bibfnamefont {P.}~\bibnamefont {Maass}},\ and\ \bibinfo
  {author} {\bibfnamefont {P.}~\bibnamefont {Tierno}},\ }\bibfield  {title}
  {\bibinfo {title} {Emergent colloidal currents across ordered and disordered
  landscapes},\ }\href {https://doi.org/10.1038/s42005-021-00722-0} {\bibfield
  {journal} {\bibinfo  {journal} {Commun. Phys.}\ }\textbf {\bibinfo {volume}
  {4}},\ \bibinfo {pages} {224} (\bibinfo {year} {2021})}\BibitemShut {NoStop}%
\bibitem [{\citenamefont {Sullivan}\ \emph {et~al.}(2002)\citenamefont
  {Sullivan}, \citenamefont {Zhao}, \citenamefont {Harrison}, \citenamefont
  {Austin}, \citenamefont {Megens}, \citenamefont {Hollingsworth},
  \citenamefont {Russel}, \citenamefont {Cheng}, \citenamefont {Mason},\ and\
  \citenamefont {Chaikin}}]{Sullivan2002}%
  \BibitemOpen
  \bibfield  {author} {\bibinfo {author} {\bibfnamefont {M.}~\bibnamefont
  {Sullivan}}, \bibinfo {author} {\bibfnamefont {K.}~\bibnamefont {Zhao}},
  \bibinfo {author} {\bibfnamefont {C.}~\bibnamefont {Harrison}}, \bibinfo
  {author} {\bibfnamefont {R.~H.}\ \bibnamefont {Austin}}, \bibinfo {author}
  {\bibfnamefont {M.}~\bibnamefont {Megens}}, \bibinfo {author} {\bibfnamefont
  {A.}~\bibnamefont {Hollingsworth}}, \bibinfo {author} {\bibfnamefont {W.~B.}\
  \bibnamefont {Russel}}, \bibinfo {author} {\bibfnamefont {Z.}~\bibnamefont
  {Cheng}}, \bibinfo {author} {\bibfnamefont {T.}~\bibnamefont {Mason}},\ and\
  \bibinfo {author} {\bibfnamefont {P.~M.}\ \bibnamefont {Chaikin}},\
  }\bibfield  {title} {\bibinfo {title} {Control of colloids with gravity,
  temperature gradients, and electric fields},\ }\href
  {https://doi.org/10.1088/0953-8984/15/1/302} {\bibfield  {journal} {\bibinfo
  {journal} {J. Phys.: Condens. Matter}\ }\textbf {\bibinfo {volume} {15}},\
  \bibinfo {pages} {S11–S18} (\bibinfo {year} {2002})}\BibitemShut {NoStop}%
\bibitem [{\citenamefont {Eckert}\ \emph {et~al.}(2021)\citenamefont {Eckert},
  \citenamefont {Schmidt},\ and\ \citenamefont {de~las Heras}}]{Eckert2021}%
  \BibitemOpen
  \bibfield  {author} {\bibinfo {author} {\bibfnamefont {T.}~\bibnamefont
  {Eckert}}, \bibinfo {author} {\bibfnamefont {M.}~\bibnamefont {Schmidt}},\
  and\ \bibinfo {author} {\bibfnamefont {D.}~\bibnamefont {de~las Heras}},\
  }\bibfield  {title} {\bibinfo {title} {Gravity-induced phase phenomena in
  plate-rod colloidal mixtures},\ }\href
  {https://doi.org/10.1038/s42005-021-00706-0} {\bibfield  {journal} {\bibinfo
  {journal} {Commun. Phys.}\ }\textbf {\bibinfo {volume} {4}},\ \bibinfo
  {pages} {202} (\bibinfo {year} {2021})}\BibitemShut {NoStop}%
\bibitem [{\citenamefont {Isele}\ \emph {et~al.}(2023)\citenamefont {Isele},
  \citenamefont {Hofmann}, \citenamefont {Erbe}, \citenamefont {Leiderer},\
  and\ \citenamefont {Nielaba}}]{Isele2023}%
  \BibitemOpen
  \bibfield  {author} {\bibinfo {author} {\bibfnamefont {M.}~\bibnamefont
  {Isele}}, \bibinfo {author} {\bibfnamefont {K.}~\bibnamefont {Hofmann}},
  \bibinfo {author} {\bibfnamefont {A.}~\bibnamefont {Erbe}}, \bibinfo {author}
  {\bibfnamefont {P.}~\bibnamefont {Leiderer}},\ and\ \bibinfo {author}
  {\bibfnamefont {P.}~\bibnamefont {Nielaba}},\ }\bibfield  {title} {\bibinfo
  {title} {Lane formation of colloidal particles driven in parallel by
  gravity},\ }\href {https://doi.org/10.1103/PhysRevE.108.034607} {\bibfield
  {journal} {\bibinfo  {journal} {Phys. Rev. E}\ }\textbf {\bibinfo {volume}
  {108}},\ \bibinfo {pages} {034607} (\bibinfo {year} {2023})}\BibitemShut
  {NoStop}%
\bibitem [{\citenamefont {Faucheux}\ \emph {et~al.}(1995)\citenamefont
  {Faucheux}, \citenamefont {Bourdieu}, \citenamefont {Kaplan},\ and\
  \citenamefont {Libchaber}}]{Faucheux1995}%
  \BibitemOpen
  \bibfield  {author} {\bibinfo {author} {\bibfnamefont {L.~P.}\ \bibnamefont
  {Faucheux}}, \bibinfo {author} {\bibfnamefont {L.~S.}\ \bibnamefont
  {Bourdieu}}, \bibinfo {author} {\bibfnamefont {P.~D.}\ \bibnamefont
  {Kaplan}},\ and\ \bibinfo {author} {\bibfnamefont {A.~J.}\ \bibnamefont
  {Libchaber}},\ }\bibfield  {title} {\bibinfo {title} {Optical thermal
  ratchet},\ }\href {https://doi.org/10.1103/physrevlett.74.1504} {\bibfield
  {journal} {\bibinfo  {journal} {Phys. Rev. Lett.}\ }\textbf {\bibinfo
  {volume} {74}},\ \bibinfo {pages} {1504} (\bibinfo {year}
  {1995})}\BibitemShut {NoStop}%
\bibitem [{\citenamefont {Reichhardt}\ and\ \citenamefont
  {Reichhardt}(2016)}]{Reichhardt2016}%
  \BibitemOpen
  \bibfield  {author} {\bibinfo {author} {\bibfnamefont {C.}~\bibnamefont
  {Reichhardt}}\ and\ \bibinfo {author} {\bibfnamefont {C.~J.~O.}\ \bibnamefont
  {Reichhardt}},\ }\bibfield  {title} {\bibinfo {title} {Depinning and
  nonequilibrium dynamic phases of particle assemblies driven over random and
  ordered substrates: a review},\ }\href
  {https://doi.org/10.1088/1361-6633/80/2/026501} {\bibfield  {journal}
  {\bibinfo  {journal} {Rep. Prog. Phys.}\ }\textbf {\bibinfo {volume} {80}},\
  \bibinfo {pages} {026501} (\bibinfo {year} {2016})}\BibitemShut {NoStop}%
\bibitem [{\citenamefont {Figueroa-Morales}\ \emph {et~al.}(2022)\citenamefont
  {Figueroa-Morales}, \citenamefont {Genkin}, \citenamefont {Sokolov},\ and\
  \citenamefont {Aranson}}]{FigueroaMorales2022}%
  \BibitemOpen
  \bibfield  {author} {\bibinfo {author} {\bibfnamefont {N.}~\bibnamefont
  {Figueroa-Morales}}, \bibinfo {author} {\bibfnamefont {M.~M.}\ \bibnamefont
  {Genkin}}, \bibinfo {author} {\bibfnamefont {A.}~\bibnamefont {Sokolov}},\
  and\ \bibinfo {author} {\bibfnamefont {I.~S.}\ \bibnamefont {Aranson}},\
  }\bibfield  {title} {\bibinfo {title} {Non-symmetric pinning of topological
  defects in living liquid crystals},\ }\href
  {https://doi.org/10.1038/s42005-022-01077-w} {\bibfield  {journal} {\bibinfo
  {journal} {Commun. Phys.}\ }\textbf {\bibinfo {volume} {5}},\ \bibinfo
  {pages} {301} (\bibinfo {year} {2022})}\BibitemShut {NoStop}%
\bibitem [{\citenamefont {Schilling}(2022)}]{Schilling2022}%
  \BibitemOpen
  \bibfield  {author} {\bibinfo {author} {\bibfnamefont {T.}~\bibnamefont
  {Schilling}},\ }\bibfield  {title} {\bibinfo {title} {Coarse-grained
  modelling out of equilibrium},\ }\href
  {https://doi.org/10.1016/j.physrep.2022.04.006} {\bibfield  {journal}
  {\bibinfo  {journal} {Phys. Rep.}\ }\textbf {\bibinfo {volume} {972}},\
  \bibinfo {pages} {1} (\bibinfo {year} {2022})}\BibitemShut {NoStop}%
\bibitem [{\citenamefont {Schmidt}\ and\ \citenamefont
  {Brader}(2013)}]{Schmidt2013}%
  \BibitemOpen
  \bibfield  {author} {\bibinfo {author} {\bibfnamefont {M.}~\bibnamefont
  {Schmidt}}\ and\ \bibinfo {author} {\bibfnamefont {J.~M.}\ \bibnamefont
  {Brader}},\ }\bibfield  {title} {\bibinfo {title} {Power functional theory
  for {B}rownian dynamics},\ }\href {https://doi.org/10.1063/1.4807586}
  {\bibfield  {journal} {\bibinfo  {journal} {J. Chem. Phys.}\ }\textbf
  {\bibinfo {volume} {138}},\ \bibinfo {pages} {214101} (\bibinfo {year}
  {2013})}\BibitemShut {NoStop}%
\bibitem [{\citenamefont {Schmidt}(2018)}]{Schmidt2018}%
  \BibitemOpen
  \bibfield  {author} {\bibinfo {author} {\bibfnamefont {M.}~\bibnamefont
  {Schmidt}},\ }\bibfield  {title} {\bibinfo {title} {Power functional theory
  for {N}ewtonian many-body dynamics},\ }\href
  {https://doi.org/10.1063/1.5008608} {\bibfield  {journal} {\bibinfo
  {journal} {J. Chem. Phys.}\ }\textbf {\bibinfo {volume} {148}},\ \bibinfo
  {pages} {044502} (\bibinfo {year} {2018})}\BibitemShut {NoStop}%
\bibitem [{\citenamefont {Schmidt}(2015)}]{Schmidt2015}%
  \BibitemOpen
  \bibfield  {author} {\bibinfo {author} {\bibfnamefont {M.}~\bibnamefont
  {Schmidt}},\ }\bibfield  {title} {\bibinfo {title} {Quantum power functional
  theory for many-body dynamics},\ }\href {https://doi.org/10.1063/1.4934881}
  {\bibfield  {journal} {\bibinfo  {journal} {J. Chem. Phys.}\ }\textbf
  {\bibinfo {volume} {143}},\ \bibinfo {pages} {174108} (\bibinfo {year}
  {2015})}\BibitemShut {NoStop}%
\bibitem [{\citenamefont {Schmidt}(2022)}]{Schmidt2022}%
  \BibitemOpen
  \bibfield  {author} {\bibinfo {author} {\bibfnamefont {M.}~\bibnamefont
  {Schmidt}},\ }\bibfield  {title} {\bibinfo {title} {Power functional theory
  for many-body dynamics},\ }\href
  {https://doi.org/10.1103/RevModPhys.94.015007} {\bibfield  {journal}
  {\bibinfo  {journal} {Rev. Mod. Phys.}\ }\textbf {\bibinfo {volume} {94}},\
  \bibinfo {pages} {015007} (\bibinfo {year} {2022})}\BibitemShut {NoStop}%
\bibitem [{\citenamefont {Evans}(1979)}]{Evans1979}%
  \BibitemOpen
  \bibfield  {author} {\bibinfo {author} {\bibfnamefont {R.}~\bibnamefont
  {Evans}},\ }\bibfield  {title} {\bibinfo {title} {The nature of the
  liquid-vapour interface and other topics in the statistical mechanics of
  non-uniform, classical fluids},\ }\href
  {https://doi.org/10.1080/00018737900101365} {\bibfield  {journal} {\bibinfo
  {journal} {Adv. Phys.}\ }\textbf {\bibinfo {volume} {28}},\ \bibinfo {pages}
  {143} (\bibinfo {year} {1979})}\BibitemShut {NoStop}%
\bibitem [{Note1()}]{Note1}%
  \BibitemOpen
  \bibinfo {note} {The one-body direct correlation functional $c_1(\protect
  \textbf {r};[\rho ])$ is related to the excess (over ideal gas) free energy
  functional $F_{\protect \rm exc}[\rho ]$ via $c_1(\protect \textbf {r};[\rho
  ])=-\delta \beta F_{\protect \rm exc}[\rho ]/\delta \rho (\protect \textbf
  {r})$ with $\beta =1/k_BT$. The internal force field is then related to the
  one-body direct correlation function via $\protect \textbf {f}_{\protect \rm
  {int}}(\protect \textbf {r};[\rho ])=k_BT\nabla c_1(\protect \textbf
  {r};[\rho ])$. Here $k_B$ is the Boltzmann constant and $T$ is (absolute)
  temperature.}\BibitemShut {Stop}%
\bibitem [{\citenamefont {Santos-Silva}\ \emph {et~al.}(2014)\citenamefont
  {Santos-Silva}, \citenamefont {Teixeira}, \citenamefont {Anquetil-Deck},\
  and\ \citenamefont {Cleaver}}]{Cleaver2014}%
  \BibitemOpen
  \bibfield  {author} {\bibinfo {author} {\bibfnamefont {T.}~\bibnamefont
  {Santos-Silva}}, \bibinfo {author} {\bibfnamefont {P.~I.~C.}\ \bibnamefont
  {Teixeira}}, \bibinfo {author} {\bibfnamefont {C.}~\bibnamefont
  {Anquetil-Deck}},\ and\ \bibinfo {author} {\bibfnamefont {D.~J.}\
  \bibnamefont {Cleaver}},\ }\bibfield  {title} {\bibinfo {title}
  {Neural-network approach to modeling liquid crystals in complex
  confinement},\ }\href {https://doi.org/10.1103/PhysRevE.89.053316} {\bibfield
   {journal} {\bibinfo  {journal} {Phys. Rev. E}\ }\textbf {\bibinfo {volume}
  {89}},\ \bibinfo {pages} {053316} (\bibinfo {year} {2014})}\BibitemShut
  {NoStop}%
\bibitem [{\citenamefont {Lin}\ and\ \citenamefont {Oettel}(2019)}]{Lin2019}%
  \BibitemOpen
  \bibfield  {author} {\bibinfo {author} {\bibfnamefont {S.-C.}\ \bibnamefont
  {Lin}}\ and\ \bibinfo {author} {\bibfnamefont {M.}~\bibnamefont {Oettel}},\
  }\bibfield  {title} {\bibinfo {title} {{A classical density functional from
  machine learning and a convolutional neural network}},\ }\href
  {https://doi.org/10.21468/SciPostPhys.6.2.025} {\bibfield  {journal}
  {\bibinfo  {journal} {SciPost Phys.}\ }\textbf {\bibinfo {volume} {6}},\
  \bibinfo {pages} {025} (\bibinfo {year} {2019})}\BibitemShut {NoStop}%
\bibitem [{\citenamefont {Lin}\ \emph {et~al.}(2020)\citenamefont {Lin},
  \citenamefont {Martius},\ and\ \citenamefont {Oettel}}]{Lin2020}%
  \BibitemOpen
  \bibfield  {author} {\bibinfo {author} {\bibfnamefont {S.-C.}\ \bibnamefont
  {Lin}}, \bibinfo {author} {\bibfnamefont {G.}~\bibnamefont {Martius}},\ and\
  \bibinfo {author} {\bibfnamefont {M.}~\bibnamefont {Oettel}},\ }\bibfield
  {title} {\bibinfo {title} {{Analytical classical density functionals from an
  equation learning network}},\ }\href {https://doi.org/10.1063/1.5135919}
  {\bibfield  {journal} {\bibinfo  {journal} {J. Chem. Phys.}\ }\textbf
  {\bibinfo {volume} {152}},\ \bibinfo {pages} {021102} (\bibinfo {year}
  {2020})}\BibitemShut {NoStop}%
\bibitem [{\citenamefont {Cats}\ \emph {et~al.}(2021)\citenamefont {Cats},
  \citenamefont {Kuipers}, \citenamefont {de~Wind}, \citenamefont {van Damme},
  \citenamefont {Coli}, \citenamefont {Dijkstra},\ and\ \citenamefont {van
  Roij}}]{Cats2021}%
  \BibitemOpen
  \bibfield  {author} {\bibinfo {author} {\bibfnamefont {P.}~\bibnamefont
  {Cats}}, \bibinfo {author} {\bibfnamefont {S.}~\bibnamefont {Kuipers}},
  \bibinfo {author} {\bibfnamefont {S.}~\bibnamefont {de~Wind}}, \bibinfo
  {author} {\bibfnamefont {R.}~\bibnamefont {van Damme}}, \bibinfo {author}
  {\bibfnamefont {G.~M.}\ \bibnamefont {Coli}}, \bibinfo {author}
  {\bibfnamefont {M.}~\bibnamefont {Dijkstra}},\ and\ \bibinfo {author}
  {\bibfnamefont {R.}~\bibnamefont {van Roij}},\ }\bibfield  {title} {\bibinfo
  {title} {{Machine-learning free-energy functionals using density profiles
  from simulations}},\ }\href {https://doi.org/10.1063/5.0042558} {\bibfield
  {journal} {\bibinfo  {journal} {APL Mater.}\ }\textbf {\bibinfo {volume}
  {9}},\ \bibinfo {pages} {031109} (\bibinfo {year} {2021})}\BibitemShut
  {NoStop}%
\bibitem [{\citenamefont {Malpica-Morales}\ \emph {et~al.}(2023)\citenamefont
  {Malpica-Morales}, \citenamefont {Yatsyshin}, \citenamefont
  {Durán-Olivencia},\ and\ \citenamefont {Kalliadasis}}]{Morales2023}%
  \BibitemOpen
  \bibfield  {author} {\bibinfo {author} {\bibfnamefont {A.}~\bibnamefont
  {Malpica-Morales}}, \bibinfo {author} {\bibfnamefont {P.}~\bibnamefont
  {Yatsyshin}}, \bibinfo {author} {\bibfnamefont {M.~A.}\ \bibnamefont
  {Durán-Olivencia}},\ and\ \bibinfo {author} {\bibfnamefont {S.}~\bibnamefont
  {Kalliadasis}},\ }\bibfield  {title} {\bibinfo {title} {{Physics-informed
  Bayesian inference of external potentials in classical density-functional
  theory}},\ }\href {https://doi.org/10.1063/5.0146920} {\bibfield  {journal}
  {\bibinfo  {journal} {J. Chem. Phys.}\ }\textbf {\bibinfo {volume} {159}},\
  \bibinfo {pages} {104109} (\bibinfo {year} {2023})}\BibitemShut {NoStop}%
\bibitem [{\citenamefont {Samm\"{u}ller}\ \emph
  {et~al.}(2023{\natexlab{a}})\citenamefont {Samm\"{u}ller}, \citenamefont
  {Hermann}, \citenamefont {de~las Heras},\ and\ \citenamefont
  {Schmidt}}]{Sammuller2023}%
  \BibitemOpen
  \bibfield  {author} {\bibinfo {author} {\bibfnamefont {F.}~\bibnamefont
  {Samm\"{u}ller}}, \bibinfo {author} {\bibfnamefont {S.}~\bibnamefont
  {Hermann}}, \bibinfo {author} {\bibfnamefont {D.}~\bibnamefont {de~las
  Heras}},\ and\ \bibinfo {author} {\bibfnamefont {M.}~\bibnamefont
  {Schmidt}},\ }\bibfield  {title} {\bibinfo {title} {Neural functional theory
  for inhomogeneous fluids: Fundamentals and applications},\ }\href
  {https://doi.org/10.1073/pnas.2312484120} {\bibfield  {journal} {\bibinfo
  {journal} {Proc. Natl. Acad. Sci.}\ }\textbf {\bibinfo {volume} {120}},\
  \bibinfo {pages} {e2312484120} (\bibinfo {year}
  {2023}{\natexlab{a}})}\BibitemShut {NoStop}%
\bibitem [{\citenamefont {Simon}\ \emph {et~al.}(2024)\citenamefont {Simon},
  \citenamefont {Weimar}, \citenamefont {Martius},\ and\ \citenamefont
  {Oettel}}]{Simon2024}%
  \BibitemOpen
  \bibfield  {author} {\bibinfo {author} {\bibfnamefont {A.}~\bibnamefont
  {Simon}}, \bibinfo {author} {\bibfnamefont {J.}~\bibnamefont {Weimar}},
  \bibinfo {author} {\bibfnamefont {G.}~\bibnamefont {Martius}},\ and\ \bibinfo
  {author} {\bibfnamefont {M.}~\bibnamefont {Oettel}},\ }\bibfield  {title}
  {\bibinfo {title} {Machine learning of a density functional for anisotropic
  patchy particles},\ }\href {https://doi.org/10.1021/acs.jctc.3c01238}
  {\bibfield  {journal} {\bibinfo  {journal} {J. Chem. Theory Comput.}\
  }\textbf {\bibinfo {volume} {20}},\ \bibinfo {pages} {1062} (\bibinfo {year}
  {2024})}\BibitemShut {NoStop}%
\bibitem [{\citenamefont {Dijkman}\ \emph {et~al.}(2024)\citenamefont
  {Dijkman}, \citenamefont {Dijkstra}, \citenamefont {van Roij}, \citenamefont
  {Welling}, \citenamefont {van~de Meent},\ and\ \citenamefont
  {Ensing}}]{dijkman2024}%
  \BibitemOpen
  \bibfield  {author} {\bibinfo {author} {\bibfnamefont {J.}~\bibnamefont
  {Dijkman}}, \bibinfo {author} {\bibfnamefont {M.}~\bibnamefont {Dijkstra}},
  \bibinfo {author} {\bibfnamefont {R.}~\bibnamefont {van Roij}}, \bibinfo
  {author} {\bibfnamefont {M.}~\bibnamefont {Welling}}, \bibinfo {author}
  {\bibfnamefont {J.-W.}\ \bibnamefont {van~de Meent}},\ and\ \bibinfo {author}
  {\bibfnamefont {B.}~\bibnamefont {Ensing}},\ }\bibfield  {title} {\bibinfo
  {title} {Learning neural free-energy functionals with pair-correlation
  matching},\ }\bibfield  {journal} {\bibinfo  {journal} {arXiv}\ }\href
  {https://doi.org/10.48550/ARXIV.2403.15007} {10.48550/ARXIV.2403.15007}
  (\bibinfo {year} {2024}),\ \Eprint {https://arxiv.org/abs/2403.15007}
  {2403.15007} \BibitemShut {NoStop}%
\bibitem [{\citenamefont {Hansen-Goos}\ and\ \citenamefont
  {Roth}(2006)}]{HansenGoos2006}%
  \BibitemOpen
  \bibfield  {author} {\bibinfo {author} {\bibfnamefont {H.}~\bibnamefont
  {Hansen-Goos}}\ and\ \bibinfo {author} {\bibfnamefont {R.}~\bibnamefont
  {Roth}},\ }\bibfield  {title} {\bibinfo {title} {Density functional theory
  for hard-sphere mixtures: the {W}hite {B}ear version mark ii},\ }\href
  {https://doi.org/10.1088/0953-8984/18/37/002} {\bibfield  {journal} {\bibinfo
   {journal} {J. Phys.: Condens. Matter}\ }\textbf {\bibinfo {volume} {18}},\
  \bibinfo {pages} {8413–8425} (\bibinfo {year} {2006})}\BibitemShut
  {NoStop}%
\bibitem [{\citenamefont {de~las Heras}\ and\ \citenamefont
  {Schmidt}(2018{\natexlab{a}})}]{delasHeras2018}%
  \BibitemOpen
  \bibfield  {author} {\bibinfo {author} {\bibfnamefont {D.}~\bibnamefont
  {de~las Heras}}\ and\ \bibinfo {author} {\bibfnamefont {M.}~\bibnamefont
  {Schmidt}},\ }\bibfield  {title} {\bibinfo {title} {Velocity gradient power
  functional for {B}rownian dynamics},\ }\href
  {https://doi.org/10.1103/PhysRevLett.120.028001} {\bibfield  {journal}
  {\bibinfo  {journal} {Phys. Rev. Lett.}\ }\textbf {\bibinfo {volume} {120}},\
  \bibinfo {pages} {028001} (\bibinfo {year} {2018}{\natexlab{a}})}\BibitemShut
  {NoStop}%
\bibitem [{\citenamefont {Stuhlm\"uller}\ \emph {et~al.}(2018)\citenamefont
  {Stuhlm\"uller}, \citenamefont {Eckert}, \citenamefont {de~las Heras},\ and\
  \citenamefont {Schmidt}}]{Stuhlmuller2018}%
  \BibitemOpen
  \bibfield  {author} {\bibinfo {author} {\bibfnamefont {N.~C.~X.}\
  \bibnamefont {Stuhlm\"uller}}, \bibinfo {author} {\bibfnamefont
  {T.}~\bibnamefont {Eckert}}, \bibinfo {author} {\bibfnamefont
  {D.}~\bibnamefont {de~las Heras}},\ and\ \bibinfo {author} {\bibfnamefont
  {M.}~\bibnamefont {Schmidt}},\ }\bibfield  {title} {\bibinfo {title}
  {Structural nonequilibrium forces in driven colloidal systems},\ }\href
  {https://doi.org/10.1103/PhysRevLett.121.098002} {\bibfield  {journal}
  {\bibinfo  {journal} {Phys. Rev. Lett.}\ }\textbf {\bibinfo {volume} {121}},\
  \bibinfo {pages} {098002} (\bibinfo {year} {2018})}\BibitemShut {NoStop}%
\bibitem [{\citenamefont {Samm\"{u}ller}\ \emph
  {et~al.}(2023{\natexlab{b}})\citenamefont {Samm\"{u}ller}, \citenamefont
  {de~las Heras},\ and\ \citenamefont {Schmidt}}]{Sammller2023}%
  \BibitemOpen
  \bibfield  {author} {\bibinfo {author} {\bibfnamefont {F.}~\bibnamefont
  {Samm\"{u}ller}}, \bibinfo {author} {\bibfnamefont {D.}~\bibnamefont {de~las
  Heras}},\ and\ \bibinfo {author} {\bibfnamefont {M.}~\bibnamefont
  {Schmidt}},\ }\bibfield  {title} {\bibinfo {title} {Inhomogeneous steady
  shear dynamics of a three-body colloidal gel former},\ }\href
  {https://doi.org/10.1063/5.0130655} {\bibfield  {journal} {\bibinfo
  {journal} {J. Chem. Phys.}\ }\textbf {\bibinfo {volume} {158}},\ \bibinfo
  {pages} {054908} (\bibinfo {year} {2023}{\natexlab{b}})}\BibitemShut
  {NoStop}%
\bibitem [{\citenamefont {de~las Heras}\ and\ \citenamefont
  {Schmidt}(2020)}]{delasHeras2020}%
  \BibitemOpen
  \bibfield  {author} {\bibinfo {author} {\bibfnamefont {D.}~\bibnamefont
  {de~las Heras}}\ and\ \bibinfo {author} {\bibfnamefont {M.}~\bibnamefont
  {Schmidt}},\ }\bibfield  {title} {\bibinfo {title} {Flow and structure in
  nonequilibrium {B}rownian many-body systems},\ }\href
  {https://doi.org/10.1103/PhysRevLett.125.018001} {\bibfield  {journal}
  {\bibinfo  {journal} {Phys. Rev. Lett.}\ }\textbf {\bibinfo {volume} {125}},\
  \bibinfo {pages} {018001} (\bibinfo {year} {2020})}\BibitemShut {NoStop}%
\bibitem [{\citenamefont {Geigenfeind}\ \emph {et~al.}(2020)\citenamefont
  {Geigenfeind}, \citenamefont {de~las Heras},\ and\ \citenamefont
  {Schmidt}}]{Geigenfeind2020}%
  \BibitemOpen
  \bibfield  {author} {\bibinfo {author} {\bibfnamefont {T.}~\bibnamefont
  {Geigenfeind}}, \bibinfo {author} {\bibfnamefont {D.}~\bibnamefont {de~las
  Heras}},\ and\ \bibinfo {author} {\bibfnamefont {M.}~\bibnamefont
  {Schmidt}},\ }\bibfield  {title} {\bibinfo {title} {Superadiabatic demixing
  in nonequilibrium colloids},\ }\href
  {https://doi.org/10.1038/s42005-020-0287-5} {\bibfield  {journal} {\bibinfo
  {journal} {Commun. Phys.}\ }\textbf {\bibinfo {volume} {3}},\ \bibinfo
  {pages} {23} (\bibinfo {year} {2020})}\BibitemShut {NoStop}%
\bibitem [{\citenamefont {Treffenst\"{a}dt}\ and\ \citenamefont
  {Schmidt}(2021)}]{Treffenstdt2021}%
  \BibitemOpen
  \bibfield  {author} {\bibinfo {author} {\bibfnamefont {L.~L.}\ \bibnamefont
  {Treffenst\"{a}dt}}\ and\ \bibinfo {author} {\bibfnamefont {M.}~\bibnamefont
  {Schmidt}},\ }\bibfield  {title} {\bibinfo {title} {Universality in driven
  and equilibrium hard sphere liquid dynamics},\ }\href
  {http://dx.doi.org/10.1103/PhysRevLett.126.058002} {\bibfield  {journal}
  {\bibinfo  {journal} {Phys. Rev. Lett.}\ }\textbf {\bibinfo {volume} {126}},\
  \bibinfo {pages} {058002} (\bibinfo {year} {2021})}\BibitemShut {NoStop}%
\bibitem [{\citenamefont {Hermann}\ \emph {et~al.}(2019)\citenamefont
  {Hermann}, \citenamefont {Krinninger}, \citenamefont {de~las Heras},\ and\
  \citenamefont {Schmidt}}]{Hermann2019}%
  \BibitemOpen
  \bibfield  {author} {\bibinfo {author} {\bibfnamefont {S.}~\bibnamefont
  {Hermann}}, \bibinfo {author} {\bibfnamefont {P.}~\bibnamefont {Krinninger}},
  \bibinfo {author} {\bibfnamefont {D.}~\bibnamefont {de~las Heras}},\ and\
  \bibinfo {author} {\bibfnamefont {M.}~\bibnamefont {Schmidt}},\ }\bibfield
  {title} {\bibinfo {title} {Phase coexistence of active {B}rownian
  particles},\ }\href {https://doi.org/10.1103/PhysRevE.100.052604} {\bibfield
  {journal} {\bibinfo  {journal} {Phys. Rev. E}\ }\textbf {\bibinfo {volume}
  {100}},\ \bibinfo {pages} {052604} (\bibinfo {year} {2019})}\BibitemShut
  {NoStop}%
\bibitem [{\citenamefont {Hermann}\ and\ \citenamefont
  {Schmidt}(2023)}]{Hermann2023}%
  \BibitemOpen
  \bibfield  {author} {\bibinfo {author} {\bibfnamefont {S.}~\bibnamefont
  {Hermann}}\ and\ \bibinfo {author} {\bibfnamefont {M.}~\bibnamefont
  {Schmidt}},\ }\bibfield  {title} {\bibinfo {title} {Active crystallization
  from power functional theory},\ }\href
  {https://doi.org/https://doi.org/10.48550/arXiv.2308.10614} {\bibfield
  {journal} {\bibinfo  {journal} {arXiv}\ ,\ \bibinfo {pages} {2308.10614}}
  (\bibinfo {year} {2023})}\BibitemShut {NoStop}%
\bibitem [{\citenamefont {de~las Heras}\ \emph {et~al.}(2023)\citenamefont
  {de~las Heras}, \citenamefont {Zimmermann}, \citenamefont {Samm\"{u}ller},
  \citenamefont {Hermann},\ and\ \citenamefont {Schmidt}}]{delasHeras2023}%
  \BibitemOpen
  \bibfield  {author} {\bibinfo {author} {\bibfnamefont {D.}~\bibnamefont
  {de~las Heras}}, \bibinfo {author} {\bibfnamefont {T.}~\bibnamefont
  {Zimmermann}}, \bibinfo {author} {\bibfnamefont {F.}~\bibnamefont
  {Samm\"{u}ller}}, \bibinfo {author} {\bibfnamefont {S.}~\bibnamefont
  {Hermann}},\ and\ \bibinfo {author} {\bibfnamefont {M.}~\bibnamefont
  {Schmidt}},\ }\bibfield  {title} {\bibinfo {title} {Perspective: How to
  overcome dynamical density functional theory},\ }\href
  {https://doi.org/10.1088/1361-648x/accb33} {\bibfield  {journal} {\bibinfo
  {journal} {J. Phys.: Condens. Matter}\ }\textbf {\bibinfo {volume} {35}},\
  \bibinfo {pages} {271501} (\bibinfo {year} {2023})}\BibitemShut {NoStop}%
\bibitem [{\citenamefont {de~las Heras}\ \emph {et~al.}(2019)\citenamefont
  {de~las Heras}, \citenamefont {Renner},\ and\ \citenamefont
  {Schmidt}}]{delasHeras2019}%
  \BibitemOpen
  \bibfield  {author} {\bibinfo {author} {\bibfnamefont {D.}~\bibnamefont
  {de~las Heras}}, \bibinfo {author} {\bibfnamefont {J.}~\bibnamefont
  {Renner}},\ and\ \bibinfo {author} {\bibfnamefont {M.}~\bibnamefont
  {Schmidt}},\ }\bibfield  {title} {\bibinfo {title} {Custom flow in overdamped
  {B}rownian dynamics},\ }\href {https://doi.org/10.1103/PhysRevE.99.023306}
  {\bibfield  {journal} {\bibinfo  {journal} {Phys. Rev. E}\ }\textbf {\bibinfo
  {volume} {99}},\ \bibinfo {pages} {023306} (\bibinfo {year}
  {2019})}\BibitemShut {NoStop}%
\bibitem [{\citenamefont {Marconi}\ and\ \citenamefont
  {Tarazona}(1999)}]{Umberto1999}%
  \BibitemOpen
  \bibfield  {author} {\bibinfo {author} {\bibfnamefont {U.~M.~B.}\
  \bibnamefont {Marconi}}\ and\ \bibinfo {author} {\bibfnamefont
  {P.}~\bibnamefont {Tarazona}},\ }\bibfield  {title} {\bibinfo {title}
  {{Dynamic density functional theory of fluids}},\ }\href
  {https://doi.org/10.1063/1.478705} {\bibfield  {journal} {\bibinfo  {journal}
  {J. Chem. Phys.}\ }\textbf {\bibinfo {volume} {110}},\ \bibinfo {pages}
  {8032} (\bibinfo {year} {1999})}\BibitemShut {NoStop}%
\bibitem [{\citenamefont {te~Vrugt}\ \emph {et~al.}(2020)\citenamefont
  {te~Vrugt}, \citenamefont {L\"{o}wen},\ and\ \citenamefont
  {Wittkowski}}]{teVrugt2020}%
  \BibitemOpen
  \bibfield  {author} {\bibinfo {author} {\bibfnamefont {M.}~\bibnamefont
  {te~Vrugt}}, \bibinfo {author} {\bibfnamefont {H.}~\bibnamefont
  {L\"{o}wen}},\ and\ \bibinfo {author} {\bibfnamefont {R.}~\bibnamefont
  {Wittkowski}},\ }\bibfield  {title} {\bibinfo {title} {Classical dynamical
  density functional theory: from fundamentals to applications},\ }\href
  {https://doi.org/10.1080/00018732.2020.1854965} {\bibfield  {journal}
  {\bibinfo  {journal} {Adv. Phys.}\ }\textbf {\bibinfo {volume} {69}},\
  \bibinfo {pages} {121–247} (\bibinfo {year} {2020})}\BibitemShut {NoStop}%
\bibitem [{\citenamefont {Frank}\ \emph {et~al.}(2003)\citenamefont {Frank},
  \citenamefont {Anderson}, \citenamefont {Weeks},\ and\ \citenamefont
  {Morris}}]{FRANK2003}%
  \BibitemOpen
  \bibfield  {author} {\bibinfo {author} {\bibfnamefont {M.}~\bibnamefont
  {Frank}}, \bibinfo {author} {\bibfnamefont {D.}~\bibnamefont {Anderson}},
  \bibinfo {author} {\bibfnamefont {E.~R.}\ \bibnamefont {Weeks}},\ and\
  \bibinfo {author} {\bibfnamefont {J.~F.}\ \bibnamefont {Morris}},\ }\bibfield
   {title} {\bibinfo {title} {Particle migration in pressure-driven flow of a
  {B}rownian suspension},\ }\href {https://doi.org/10.1017/s0022112003006001}
  {\bibfield  {journal} {\bibinfo  {journal} {J. Fluid Mech.}\ }\textbf
  {\bibinfo {volume} {493}},\ \bibinfo {pages} {363–378} (\bibinfo {year}
  {2003})}\BibitemShut {NoStop}%
\bibitem [{\citenamefont {Leighton}\ and\ \citenamefont
  {Acrivos}(1987)}]{Leighton1987}%
  \BibitemOpen
  \bibfield  {author} {\bibinfo {author} {\bibfnamefont {D.}~\bibnamefont
  {Leighton}}\ and\ \bibinfo {author} {\bibfnamefont {A.}~\bibnamefont
  {Acrivos}},\ }\bibfield  {title} {\bibinfo {title} {The shear-induced
  migration of particles in concentrated suspensions},\ }\href
  {https://doi.org/10.1017/s0022112087002155} {\bibfield  {journal} {\bibinfo
  {journal} {J. Fluid Mech.}\ }\textbf {\bibinfo {volume} {181}},\ \bibinfo
  {pages} {415} (\bibinfo {year} {1987})}\BibitemShut {NoStop}%
\bibitem [{\citenamefont {Dzubiella}\ \emph {et~al.}(2002)\citenamefont
  {Dzubiella}, \citenamefont {Hoffmann},\ and\ \citenamefont
  {L\"owen}}]{Dzubiella2002}%
  \BibitemOpen
  \bibfield  {author} {\bibinfo {author} {\bibfnamefont {J.}~\bibnamefont
  {Dzubiella}}, \bibinfo {author} {\bibfnamefont {G.~P.}\ \bibnamefont
  {Hoffmann}},\ and\ \bibinfo {author} {\bibfnamefont {H.}~\bibnamefont
  {L\"owen}},\ }\bibfield  {title} {\bibinfo {title} {Lane formation in
  colloidal mixtures driven by an external field},\ }\href
  {https://doi.org/10.1103/PhysRevE.65.021402} {\bibfield  {journal} {\bibinfo
  {journal} {Phys. Rev. E}\ }\textbf {\bibinfo {volume} {65}},\ \bibinfo
  {pages} {021402} (\bibinfo {year} {2002})}\BibitemShut {NoStop}%
\bibitem [{\citenamefont {Samm\"{u}ller}\ and\ \citenamefont
  {Schmidt}(2021)}]{Sammller2021}%
  \BibitemOpen
  \bibfield  {author} {\bibinfo {author} {\bibfnamefont {F.}~\bibnamefont
  {Samm\"{u}ller}}\ and\ \bibinfo {author} {\bibfnamefont {M.}~\bibnamefont
  {Schmidt}},\ }\bibfield  {title} {\bibinfo {title} {Adaptive {B}rownian
  dynamics},\ }\href {https://doi.org/10.1063/5.0062396} {\bibfield  {journal}
  {\bibinfo  {journal} {J. Chem. Phys.}\ }\textbf {\bibinfo {volume} {155}},\
  \bibinfo {pages} {134107} (\bibinfo {year} {2021})}\BibitemShut {NoStop}%
\bibitem [{\citenamefont {Fortini}\ \emph {et~al.}(2014)\citenamefont
  {Fortini}, \citenamefont {de~las Heras}, \citenamefont {Brader},\ and\
  \citenamefont {Schmidt}}]{Fortini2014}%
  \BibitemOpen
  \bibfield  {author} {\bibinfo {author} {\bibfnamefont {A.}~\bibnamefont
  {Fortini}}, \bibinfo {author} {\bibfnamefont {D.}~\bibnamefont {de~las
  Heras}}, \bibinfo {author} {\bibfnamefont {J.~M.}\ \bibnamefont {Brader}},\
  and\ \bibinfo {author} {\bibfnamefont {M.}~\bibnamefont {Schmidt}},\
  }\bibfield  {title} {\bibinfo {title} {Superadiabatic forces in {B}rownian
  many-body dynamics},\ }\href {https://doi.org/10.1103/PhysRevLett.113.167801}
  {\bibfield  {journal} {\bibinfo  {journal} {Phys. Rev. Lett.}\ }\textbf
  {\bibinfo {volume} {113}},\ \bibinfo {pages} {167801} (\bibinfo {year}
  {2014})}\BibitemShut {NoStop}%
\bibitem [{\citenamefont {Cohen}\ and\ \citenamefont
  {Welling}(2016)}]{Cohen2016}%
  \BibitemOpen
  \bibfield  {author} {\bibinfo {author} {\bibfnamefont {T.}~\bibnamefont
  {Cohen}}\ and\ \bibinfo {author} {\bibfnamefont {M.}~\bibnamefont
  {Welling}},\ }\bibfield  {title} {\bibinfo {title} {Group equivariant
  convolutional networks},\ }\href
  {http://proceedings.mlr.press/v48/cohenc16.html} {\bibfield  {journal}
  {\bibinfo  {journal} {PMLR}\ }\textbf {\bibinfo {volume} {48}},\ \bibinfo
  {pages} {2990} (\bibinfo {year} {2016})}\BibitemShut {NoStop}%
\bibitem [{\citenamefont {Karniadakis}\ \emph {et~al.}(2021)\citenamefont
  {Karniadakis}, \citenamefont {Kevrekidis}, \citenamefont {Lu}, \citenamefont
  {Perdikaris}, \citenamefont {Wang},\ and\ \citenamefont
  {Yang}}]{Karniadakis2021}%
  \BibitemOpen
  \bibfield  {author} {\bibinfo {author} {\bibfnamefont {G.~E.}\ \bibnamefont
  {Karniadakis}}, \bibinfo {author} {\bibfnamefont {I.~G.}\ \bibnamefont
  {Kevrekidis}}, \bibinfo {author} {\bibfnamefont {L.}~\bibnamefont {Lu}},
  \bibinfo {author} {\bibfnamefont {P.}~\bibnamefont {Perdikaris}}, \bibinfo
  {author} {\bibfnamefont {S.}~\bibnamefont {Wang}},\ and\ \bibinfo {author}
  {\bibfnamefont {L.}~\bibnamefont {Yang}},\ }\bibfield  {title} {\bibinfo
  {title} {Physics-informed machine learning},\ }\href
  {https://doi.org/10.1038/s42254-021-00314-5} {\bibfield  {journal} {\bibinfo
  {journal} {Nat. Rev. Phys.}\ }\textbf {\bibinfo {volume} {3}},\ \bibinfo
  {pages} {422–440} (\bibinfo {year} {2021})}\BibitemShut {NoStop}%
\bibitem [{\citenamefont {Renner}\ \emph {et~al.}(2021)\citenamefont {Renner},
  \citenamefont {Schmidt},\ and\ \citenamefont {de~las Heras}}]{Renner2021}%
  \BibitemOpen
  \bibfield  {author} {\bibinfo {author} {\bibfnamefont {J.}~\bibnamefont
  {Renner}}, \bibinfo {author} {\bibfnamefont {M.}~\bibnamefont {Schmidt}},\
  and\ \bibinfo {author} {\bibfnamefont {D.}~\bibnamefont {de~las Heras}},\
  }\bibfield  {title} {\bibinfo {title} {Custom flow in molecular dynamics},\
  }\href {https://doi.org/10.1103/PhysRevResearch.3.013281} {\bibfield
  {journal} {\bibinfo  {journal} {Phys. Rev. Res.}\ }\textbf {\bibinfo {volume}
  {3}},\ \bibinfo {pages} {013281} (\bibinfo {year} {2021})}\BibitemShut
  {NoStop}%
\bibitem [{\citenamefont {Renner}\ \emph {et~al.}(2022)\citenamefont {Renner},
  \citenamefont {Schmidt},\ and\ \citenamefont {de~las Heras}}]{Renner2022}%
  \BibitemOpen
  \bibfield  {author} {\bibinfo {author} {\bibfnamefont {J.}~\bibnamefont
  {Renner}}, \bibinfo {author} {\bibfnamefont {M.}~\bibnamefont {Schmidt}},\
  and\ \bibinfo {author} {\bibfnamefont {D.}~\bibnamefont {de~las Heras}},\
  }\bibfield  {title} {\bibinfo {title} {Shear and bulk acceleration
  viscosities in simple fluids},\ }\href
  {https://doi.org/10.1103/PhysRevLett.128.094502} {\bibfield  {journal}
  {\bibinfo  {journal} {Phys. Rev. Lett.}\ }\textbf {\bibinfo {volume} {128}},\
  \bibinfo {pages} {094502} (\bibinfo {year} {2022})}\BibitemShut {NoStop}%
\bibitem [{\citenamefont {Miskin}\ \emph {et~al.}(2015)\citenamefont {Miskin},
  \citenamefont {Khaira}, \citenamefont {de~Pablo},\ and\ \citenamefont
  {Jaeger}}]{Miskin2015}%
  \BibitemOpen
  \bibfield  {author} {\bibinfo {author} {\bibfnamefont {M.~Z.}\ \bibnamefont
  {Miskin}}, \bibinfo {author} {\bibfnamefont {G.}~\bibnamefont {Khaira}},
  \bibinfo {author} {\bibfnamefont {J.~J.}\ \bibnamefont {de~Pablo}},\ and\
  \bibinfo {author} {\bibfnamefont {H.~M.}\ \bibnamefont {Jaeger}},\ }\bibfield
   {title} {\bibinfo {title} {Turning statistical physics models into materials
  design engines},\ }\href {https://doi.org/10.1073/pnas.1509316112} {\bibfield
   {journal} {\bibinfo  {journal} {Proc. Natl. Acad. Sci.}\ }\textbf {\bibinfo
  {volume} {113}},\ \bibinfo {pages} {34} (\bibinfo {year} {2015})}\BibitemShut
  {NoStop}%
\bibitem [{\citenamefont {Sherman}\ \emph {et~al.}(2020)\citenamefont
  {Sherman}, \citenamefont {Howard}, \citenamefont {Lindquist}, \citenamefont
  {Jadrich},\ and\ \citenamefont {Truskett}}]{Sherman2020}%
  \BibitemOpen
  \bibfield  {author} {\bibinfo {author} {\bibfnamefont {Z.~M.}\ \bibnamefont
  {Sherman}}, \bibinfo {author} {\bibfnamefont {M.~P.}\ \bibnamefont {Howard}},
  \bibinfo {author} {\bibfnamefont {B.~A.}\ \bibnamefont {Lindquist}}, \bibinfo
  {author} {\bibfnamefont {R.~B.}\ \bibnamefont {Jadrich}},\ and\ \bibinfo
  {author} {\bibfnamefont {T.~M.}\ \bibnamefont {Truskett}},\ }\bibfield
  {title} {\bibinfo {title} {Inverse methods for design of soft materials},\
  }\href {https://doi.org/10.1063/1.5145177} {\bibfield  {journal} {\bibinfo
  {journal} {J. Chem. Phys.}\ }\textbf {\bibinfo {volume} {152}},\ \bibinfo
  {pages} {140902} (\bibinfo {year} {2020})}\BibitemShut {NoStop}%
\bibitem [{\citenamefont {Coli}\ \emph {et~al.}(2022)\citenamefont {Coli},
  \citenamefont {Boattini}, \citenamefont {Filion},\ and\ \citenamefont
  {Dijkstra}}]{Coli2022}%
  \BibitemOpen
  \bibfield  {author} {\bibinfo {author} {\bibfnamefont {G.~M.}\ \bibnamefont
  {Coli}}, \bibinfo {author} {\bibfnamefont {E.}~\bibnamefont {Boattini}},
  \bibinfo {author} {\bibfnamefont {L.}~\bibnamefont {Filion}},\ and\ \bibinfo
  {author} {\bibfnamefont {M.}~\bibnamefont {Dijkstra}},\ }\bibfield  {title}
  {\bibinfo {title} {Inverse design of soft materials via a deep
  learning–based evolutionary strategy},\ }\href
  {https://doi.org/10.1126/sciadv.abj6731} {\bibfield  {journal} {\bibinfo
  {journal} {Sci. Adv.}\ }\textbf {\bibinfo {volume} {8}},\ \bibinfo {pages}
  {eabj6731} (\bibinfo {year} {2022})}\BibitemShut {NoStop}%
\bibitem [{\citenamefont {Obukhov}(1983)}]{Obukhov1983}%
  \BibitemOpen
  \bibfield  {author} {\bibinfo {author} {\bibfnamefont {A.~M.}\ \bibnamefont
  {Obukhov}},\ }\bibfield  {title} {\bibinfo {title} {Kolmogorov flow and
  laboratory simulation of it},\ }\href
  {https://doi.org/10.1070/rm1983v038n04abeh004207} {\bibfield  {journal}
  {\bibinfo  {journal} {Russ. Math. Surv.}\ }\textbf {\bibinfo {volume} {38}},\
  \bibinfo {pages} {113–126} (\bibinfo {year} {1983})}\BibitemShut {NoStop}%
\bibitem [{\citenamefont {Jahreis}\ and\ \citenamefont
  {Schmidt}(2024)}]{nikolai}%
  \BibitemOpen
  \bibfield  {author} {\bibinfo {author} {\bibfnamefont {N.}~\bibnamefont
  {Jahreis}}\ and\ \bibinfo {author} {\bibfnamefont {M.}~\bibnamefont
  {Schmidt}},\ }\bibfield  {title} {\bibinfo {title} {title},\ }\href@noop {}
  {\bibfield  {journal} {\bibinfo  {journal} {in preparation}\ } (\bibinfo
  {year} {2024})}\BibitemShut {NoStop}%
\bibitem [{\citenamefont {Hermann}\ and\ \citenamefont
  {Schmidt}(2021)}]{Hermann2021}%
  \BibitemOpen
  \bibfield  {author} {\bibinfo {author} {\bibfnamefont {S.}~\bibnamefont
  {Hermann}}\ and\ \bibinfo {author} {\bibfnamefont {M.}~\bibnamefont
  {Schmidt}},\ }\bibfield  {title} {\bibinfo {title} {Noether’s theorem in
  statistical mechanics},\ }\href {https://doi.org/10.1038/s42005-021-00669-2}
  {\bibfield  {journal} {\bibinfo  {journal} {Commun. Phys.}\ }\textbf
  {\bibinfo {volume} {4}},\ \bibinfo {pages} {176} (\bibinfo {year}
  {2021})}\BibitemShut {NoStop}%
\bibitem [{\citenamefont {Ling}\ \emph {et~al.}(2016)\citenamefont {Ling},
  \citenamefont {Kurzawski},\ and\ \citenamefont {Templeton}}]{Ling2016}%
  \BibitemOpen
  \bibfield  {author} {\bibinfo {author} {\bibfnamefont {J.}~\bibnamefont
  {Ling}}, \bibinfo {author} {\bibfnamefont {A.}~\bibnamefont {Kurzawski}},\
  and\ \bibinfo {author} {\bibfnamefont {J.}~\bibnamefont {Templeton}},\
  }\bibfield  {title} {\bibinfo {title} {Reynolds averaged turbulence modelling
  using deep neural networks with embedded invariance},\ }\href
  {https://doi.org/10.1017/jfm.2016.615} {\bibfield  {journal} {\bibinfo
  {journal} {J. Fluid Mech.}\ }\textbf {\bibinfo {volume} {807}},\ \bibinfo
  {pages} {155–166} (\bibinfo {year} {2016})}\BibitemShut {NoStop}%
\bibitem [{\citenamefont {Treffenst\"{a}dt}\ and\ \citenamefont
  {Schmidt}(2020)}]{Treffenstdt2020}%
  \BibitemOpen
  \bibfield  {author} {\bibinfo {author} {\bibfnamefont {L.~L.}\ \bibnamefont
  {Treffenst\"{a}dt}}\ and\ \bibinfo {author} {\bibfnamefont {M.}~\bibnamefont
  {Schmidt}},\ }\bibfield  {title} {\bibinfo {title} {Memory-induced motion
  reversal in brownian liquids},\ }\href {https://doi.org/10.1039/c9sm02005e}
  {\bibfield  {journal} {\bibinfo  {journal} {Soft Matter}\ }\textbf {\bibinfo
  {volume} {16}},\ \bibinfo {pages} {1518–1526} (\bibinfo {year}
  {2020})}\BibitemShut {NoStop}%
\bibitem [{\citenamefont {Mederos}\ \emph {et~al.}(2014)\citenamefont
  {Mederos}, \citenamefont {Velasco},\ and\ \citenamefont
  {Mart{\'{\i}}nez-Rat{\'{o}}n}}]{Mederos2014}%
  \BibitemOpen
  \bibfield  {author} {\bibinfo {author} {\bibfnamefont {L.}~\bibnamefont
  {Mederos}}, \bibinfo {author} {\bibfnamefont {E.}~\bibnamefont {Velasco}},\
  and\ \bibinfo {author} {\bibfnamefont {Y.}~\bibnamefont
  {Mart{\'{\i}}nez-Rat{\'{o}}n}},\ }\bibfield  {title} {\bibinfo {title}
  {Hard-body models of bulk liquid crystals},\ }\href
  {https://doi.org/10.1088/0953-8984/26/46/463101} {\bibfield  {journal}
  {\bibinfo  {journal} {J. Phys.: Condens. Matter}\ }\textbf {\bibinfo {volume}
  {26}},\ \bibinfo {pages} {463101} (\bibinfo {year} {2014})}\BibitemShut
  {NoStop}%
\bibitem [{\citenamefont {Rex}\ \emph {et~al.}(2007)\citenamefont {Rex},
  \citenamefont {Wensink},\ and\ \citenamefont {L\"owen}}]{Wensink2007}%
  \BibitemOpen
  \bibfield  {author} {\bibinfo {author} {\bibfnamefont {M.}~\bibnamefont
  {Rex}}, \bibinfo {author} {\bibfnamefont {H.~H.}\ \bibnamefont {Wensink}},\
  and\ \bibinfo {author} {\bibfnamefont {H.}~\bibnamefont {L\"owen}},\
  }\bibfield  {title} {\bibinfo {title} {Dynamical density functional theory
  for anisotropic colloidal particles},\ }\href
  {https://doi.org/10.1103/PhysRevE.76.021403} {\bibfield  {journal} {\bibinfo
  {journal} {Phys. Rev. E}\ }\textbf {\bibinfo {volume} {76}},\ \bibinfo
  {pages} {021403} (\bibinfo {year} {2007})}\BibitemShut {NoStop}%
\bibitem [{\citenamefont {Renner}\ \emph {et~al.}(2023)\citenamefont {Renner},
  \citenamefont {Schmidt},\ and\ \citenamefont {de~las Heras}}]{Renner2023}%
  \BibitemOpen
  \bibfield  {author} {\bibinfo {author} {\bibfnamefont {J.}~\bibnamefont
  {Renner}}, \bibinfo {author} {\bibfnamefont {M.}~\bibnamefont {Schmidt}},\
  and\ \bibinfo {author} {\bibfnamefont {D.}~\bibnamefont {de~las Heras}},\
  }\bibfield  {title} {\bibinfo {title} {Reduced-variance orientational
  distribution functions from torque sampling},\ }\href
  {https://doi.org/10.1088/1361-648x/acc522} {\bibfield  {journal} {\bibinfo
  {journal} {J. Phys.: Condens. Matter}\ }\textbf {\bibinfo {volume} {35}},\
  \bibinfo {pages} {235901} (\bibinfo {year} {2023})}\BibitemShut {NoStop}%
\bibitem [{\citenamefont {Chacón}\ \emph {et~al.}(2006)\citenamefont
  {Chacón}, \citenamefont {Tarazona},\ and\ \citenamefont
  {Alejandre}}]{Chacn2006}%
  \BibitemOpen
  \bibfield  {author} {\bibinfo {author} {\bibfnamefont {E.}~\bibnamefont
  {Chacón}}, \bibinfo {author} {\bibfnamefont {P.}~\bibnamefont {Tarazona}},\
  and\ \bibinfo {author} {\bibfnamefont {J.}~\bibnamefont {Alejandre}},\
  }\bibfield  {title} {\bibinfo {title} {The intrinsic structure of the water
  surface},\ }\href {https://doi.org/10.1063/1.2209681} {\bibfield  {journal}
  {\bibinfo  {journal} {J. Chem. Phys.}\ }\textbf {\bibinfo {volume} {125}},\
  \bibinfo {pages} {014709} (\bibinfo {year} {2006})}\BibitemShut {NoStop}%
\bibitem [{\citenamefont {Duque}\ \emph {et~al.}(2008)\citenamefont {Duque},
  \citenamefont {Tarazona},\ and\ \citenamefont {Chacón}}]{Duque2008}%
  \BibitemOpen
  \bibfield  {author} {\bibinfo {author} {\bibfnamefont {D.}~\bibnamefont
  {Duque}}, \bibinfo {author} {\bibfnamefont {P.}~\bibnamefont {Tarazona}},\
  and\ \bibinfo {author} {\bibfnamefont {E.}~\bibnamefont {Chacón}},\
  }\bibfield  {title} {\bibinfo {title} {Diffusion at the liquid-vapor
  interface},\ }\href {https://doi.org/10.1063/1.2841128} {\bibfield  {journal}
  {\bibinfo  {journal} {J. Chem. Phys.}\ }\textbf {\bibinfo {volume} {128}},\
  \bibinfo {pages} {134704} (\bibinfo {year} {2008})}\BibitemShut {NoStop}%
\bibitem [{\citenamefont {Ogawa}\ \emph {et~al.}(2019)\citenamefont {Ogawa},
  \citenamefont {Oga}, \citenamefont {Kusudo}, \citenamefont {Yamaguchi},
  \citenamefont {Omori}, \citenamefont {Merabia},\ and\ \citenamefont
  {Joly}}]{Ogawa2019}%
  \BibitemOpen
  \bibfield  {author} {\bibinfo {author} {\bibfnamefont {K.}~\bibnamefont
  {Ogawa}}, \bibinfo {author} {\bibfnamefont {H.}~\bibnamefont {Oga}}, \bibinfo
  {author} {\bibfnamefont {H.}~\bibnamefont {Kusudo}}, \bibinfo {author}
  {\bibfnamefont {Y.}~\bibnamefont {Yamaguchi}}, \bibinfo {author}
  {\bibfnamefont {T.}~\bibnamefont {Omori}}, \bibinfo {author} {\bibfnamefont
  {S.}~\bibnamefont {Merabia}},\ and\ \bibinfo {author} {\bibfnamefont
  {L.}~\bibnamefont {Joly}},\ }\bibfield  {title} {\bibinfo {title} {Large
  effect of lateral box size in molecular dynamics simulations of liquid-solid
  friction},\ }\href {https://doi.org/10.1103/PhysRevE.100.023101} {\bibfield
  {journal} {\bibinfo  {journal} {Phys. Rev. E}\ }\textbf {\bibinfo {volume}
  {100}},\ \bibinfo {pages} {023101} (\bibinfo {year} {2019})}\BibitemShut
  {NoStop}%
\bibitem [{\citenamefont {Borgis}\ \emph {et~al.}(2013)\citenamefont {Borgis},
  \citenamefont {Assaraf}, \citenamefont {Rotenberg},\ and\ \citenamefont
  {Vuilleumier}}]{Borgis2013}%
  \BibitemOpen
  \bibfield  {author} {\bibinfo {author} {\bibfnamefont {D.}~\bibnamefont
  {Borgis}}, \bibinfo {author} {\bibfnamefont {R.}~\bibnamefont {Assaraf}},
  \bibinfo {author} {\bibfnamefont {B.}~\bibnamefont {Rotenberg}},\ and\
  \bibinfo {author} {\bibfnamefont {R.}~\bibnamefont {Vuilleumier}},\
  }\bibfield  {title} {\bibinfo {title} {Computation of pair distribution
  functions and three-dimensional densities with a reduced variance
  principle},\ }\href {https://doi.org/10.1080/00268976.2013.838316} {\bibfield
   {journal} {\bibinfo  {journal} {Mol. Phys.}\ }\textbf {\bibinfo {volume}
  {111}},\ \bibinfo {pages} {3486} (\bibinfo {year} {2013})}\BibitemShut
  {NoStop}%
\bibitem [{\citenamefont {de~las Heras}\ and\ \citenamefont
  {Schmidt}(2018{\natexlab{b}})}]{Heras2018a}%
  \BibitemOpen
  \bibfield  {author} {\bibinfo {author} {\bibfnamefont {D.}~\bibnamefont
  {de~las Heras}}\ and\ \bibinfo {author} {\bibfnamefont {M.}~\bibnamefont
  {Schmidt}},\ }\bibfield  {title} {\bibinfo {title} {Better than counting:
  Density profiles from force sampling},\ }\href
  {https://doi.org/10.1103/physrevlett.120.218001} {\bibfield  {journal}
  {\bibinfo  {journal} {Phys. Rev. Lett.}\ }\textbf {\bibinfo {volume} {120}},\
  \bibinfo {pages} {218001} (\bibinfo {year} {2018}{\natexlab{b}})}\BibitemShut
  {NoStop}%
\bibitem [{\citenamefont {Schultz}\ \emph {et~al.}(2016)\citenamefont
  {Schultz}, \citenamefont {Moustafa}, \citenamefont {Lin}, \citenamefont
  {Weinstein},\ and\ \citenamefont {Kofke}}]{Schultz2016}%
  \BibitemOpen
  \bibfield  {author} {\bibinfo {author} {\bibfnamefont {A.~J.}\ \bibnamefont
  {Schultz}}, \bibinfo {author} {\bibfnamefont {S.~G.}\ \bibnamefont
  {Moustafa}}, \bibinfo {author} {\bibfnamefont {W.}~\bibnamefont {Lin}},
  \bibinfo {author} {\bibfnamefont {S.~J.}\ \bibnamefont {Weinstein}},\ and\
  \bibinfo {author} {\bibfnamefont {D.~A.}\ \bibnamefont {Kofke}},\ }\bibfield
  {title} {\bibinfo {title} {Reformulation of ensemble averages via coordinate
  mapping},\ }\href {https://doi.org/10.1021/acs.jctc.6b00018} {\bibfield
  {journal} {\bibinfo  {journal} {J. Chem. Theory Comput.}\ }\textbf {\bibinfo
  {volume} {12}},\ \bibinfo {pages} {1491} (\bibinfo {year}
  {2016})}\BibitemShut {NoStop}%
\bibitem [{\citenamefont {Rotenberg}(2020)}]{Rotenberg2020}%
  \BibitemOpen
  \bibfield  {author} {\bibinfo {author} {\bibfnamefont {B.}~\bibnamefont
  {Rotenberg}},\ }\bibfield  {title} {\bibinfo {title} {Use the force!
  {R}educed variance estimators for densities, radial distribution functions,
  and local mobilities in molecular simulations},\ }\href
  {https://doi.org/10.1063/5.0029113} {\bibfield  {journal} {\bibinfo
  {journal} {J. Chem. Phys.}\ }\textbf {\bibinfo {volume} {153}},\ \bibinfo
  {pages} {150902} (\bibinfo {year} {2020})}\BibitemShut {NoStop}%
\bibitem [{\citenamefont {Margossian}(2019)}]{Margossian2019}%
  \BibitemOpen
  \bibfield  {author} {\bibinfo {author} {\bibfnamefont {C.~C.}\ \bibnamefont
  {Margossian}},\ }\bibfield  {title} {\bibinfo {title} {A review of automatic
  differentiation and its efficient implementation},\ }\href
  {https://doi.org/10.1002/widm.1305} {\bibfield  {journal} {\bibinfo
  {journal} {Data Min. Knowl. Discov.}\ }\textbf {\bibinfo {volume} {9}},\
  \bibinfo {pages} {e1305} (\bibinfo {year} {2019})}\BibitemShut {NoStop}%
\bibitem [{\citenamefont {Brader}\ and\ \citenamefont
  {Schmidt}(2015)}]{Brader2015}%
  \BibitemOpen
  \bibfield  {author} {\bibinfo {author} {\bibfnamefont {J.~M.}\ \bibnamefont
  {Brader}}\ and\ \bibinfo {author} {\bibfnamefont {M.}~\bibnamefont
  {Schmidt}},\ }\bibfield  {title} {\bibinfo {title} {Free power dissipation
  from functional line integration},\ }\href
  {https://doi.org/10.1080/00268976.2015.1042086} {\bibfield  {journal}
  {\bibinfo  {journal} {Mol. Phys.}\ }\textbf {\bibinfo {volume} {113}},\
  \bibinfo {pages} {2873–2880} (\bibinfo {year} {2015})}\BibitemShut
  {NoStop}%
\bibitem [{\citenamefont {Lesnicki}\ \emph {et~al.}(2016)\citenamefont
  {Lesnicki}, \citenamefont {Vuilleumier}, \citenamefont {Carof},\ and\
  \citenamefont {Rotenberg}}]{Lesnicki2016}%
  \BibitemOpen
  \bibfield  {author} {\bibinfo {author} {\bibfnamefont {D.}~\bibnamefont
  {Lesnicki}}, \bibinfo {author} {\bibfnamefont {R.}~\bibnamefont
  {Vuilleumier}}, \bibinfo {author} {\bibfnamefont {A.}~\bibnamefont {Carof}},\
  and\ \bibinfo {author} {\bibfnamefont {B.}~\bibnamefont {Rotenberg}},\
  }\bibfield  {title} {\bibinfo {title} {Molecular hydrodynamics from memory
  kernels},\ }\href {https://doi.org/10.1103/PhysRevLett.116.147804} {\bibfield
   {journal} {\bibinfo  {journal} {Phys. Rev. Lett.}\ }\textbf {\bibinfo
  {volume} {116}},\ \bibinfo {pages} {147804} (\bibinfo {year}
  {2016})}\BibitemShut {NoStop}%
\bibitem [{\citenamefont {Jung}\ \emph {et~al.}(2017)\citenamefont {Jung},
  \citenamefont {Hanke},\ and\ \citenamefont {Schmid}}]{Jung2017}%
  \BibitemOpen
  \bibfield  {author} {\bibinfo {author} {\bibfnamefont {G.}~\bibnamefont
  {Jung}}, \bibinfo {author} {\bibfnamefont {M.}~\bibnamefont {Hanke}},\ and\
  \bibinfo {author} {\bibfnamefont {F.}~\bibnamefont {Schmid}},\ }\bibfield
  {title} {\bibinfo {title} {Iterative reconstruction of memory kernels},\
  }\href {https://doi.org/10.1021/acs.jctc.7b00274} {\bibfield  {journal}
  {\bibinfo  {journal} {J. Chem. Theory Comput.}\ }\textbf {\bibinfo {volume}
  {13}},\ \bibinfo {pages} {2481} (\bibinfo {year} {2017})}\BibitemShut
  {NoStop}%
\bibitem [{\citenamefont {Daldrop}\ \emph {et~al.}(2017)\citenamefont
  {Daldrop}, \citenamefont {Kowalik},\ and\ \citenamefont
  {Netz}}]{Daldrop2017}%
  \BibitemOpen
  \bibfield  {author} {\bibinfo {author} {\bibfnamefont {J.~O.}\ \bibnamefont
  {Daldrop}}, \bibinfo {author} {\bibfnamefont {B.~G.}\ \bibnamefont
  {Kowalik}},\ and\ \bibinfo {author} {\bibfnamefont {R.~R.}\ \bibnamefont
  {Netz}},\ }\bibfield  {title} {\bibinfo {title} {External potential modifies
  friction of molecular solutes in water},\ }\href
  {https://doi.org/10.1103/PhysRevX.7.041065} {\bibfield  {journal} {\bibinfo
  {journal} {Phys. Rev. X}\ }\textbf {\bibinfo {volume} {7}},\ \bibinfo {pages}
  {041065} (\bibinfo {year} {2017})}\BibitemShut {NoStop}%
\bibitem [{\citenamefont {Meyer}\ \emph {et~al.}(2020)\citenamefont {Meyer},
  \citenamefont {Pelagejcev},\ and\ \citenamefont {Schilling}}]{Meyer2020}%
  \BibitemOpen
  \bibfield  {author} {\bibinfo {author} {\bibfnamefont {H.}~\bibnamefont
  {Meyer}}, \bibinfo {author} {\bibfnamefont {P.}~\bibnamefont {Pelagejcev}},\
  and\ \bibinfo {author} {\bibfnamefont {T.}~\bibnamefont {Schilling}},\
  }\bibfield  {title} {\bibinfo {title} {Non-{M}arkovian out-of-equilibrium
  dynamics: A general numerical procedure to construct time-dependent memory
  kernels for coarse-grained observables},\ }\href
  {https://doi.org/10.1209/0295-5075/128/40001} {\bibfield  {journal} {\bibinfo
   {journal} {EPL}\ }\textbf {\bibinfo {volume} {128}},\ \bibinfo {pages}
  {40001} (\bibinfo {year} {2020})}\BibitemShut {NoStop}%
\bibitem [{\citenamefont {Br\"{u}tting}\ \emph {et~al.}(2019)\citenamefont
  {Br\"{u}tting}, \citenamefont {Trepl}, \citenamefont {de~las Heras},\ and\
  \citenamefont {Schmidt}}]{Brtting2019}%
  \BibitemOpen
  \bibfield  {author} {\bibinfo {author} {\bibfnamefont {M.}~\bibnamefont
  {Br\"{u}tting}}, \bibinfo {author} {\bibfnamefont {T.}~\bibnamefont {Trepl}},
  \bibinfo {author} {\bibfnamefont {D.}~\bibnamefont {de~las Heras}},\ and\
  \bibinfo {author} {\bibfnamefont {M.}~\bibnamefont {Schmidt}},\ }\bibfield
  {title} {\bibinfo {title} {Superadiabatic forces via the acceleration
  gradient in quantum many-body dynamics},\ }\href
  {https://doi.org/10.3390/molecules24203660} {\bibfield  {journal} {\bibinfo
  {journal} {Molecules}\ }\textbf {\bibinfo {volume} {24}},\ \bibinfo {pages}
  {3660} (\bibinfo {year} {2019})}\BibitemShut {NoStop}%
\bibitem [{\citenamefont {Sch\"{a}ffner}\ \emph {et~al.}(2020)\citenamefont
  {Sch\"{a}ffner}, \citenamefont {Preuschoff}, \citenamefont {Ristok},
  \citenamefont {Brozio}, \citenamefont {Schlosser}, \citenamefont {Giessen},\
  and\ \citenamefont {Birkl}}]{Schffner2020}%
  \BibitemOpen
  \bibfield  {author} {\bibinfo {author} {\bibfnamefont {D.}~\bibnamefont
  {Sch\"{a}ffner}}, \bibinfo {author} {\bibfnamefont {T.}~\bibnamefont
  {Preuschoff}}, \bibinfo {author} {\bibfnamefont {S.}~\bibnamefont {Ristok}},
  \bibinfo {author} {\bibfnamefont {L.}~\bibnamefont {Brozio}}, \bibinfo
  {author} {\bibfnamefont {M.}~\bibnamefont {Schlosser}}, \bibinfo {author}
  {\bibfnamefont {H.}~\bibnamefont {Giessen}},\ and\ \bibinfo {author}
  {\bibfnamefont {G.}~\bibnamefont {Birkl}},\ }\bibfield  {title} {\bibinfo
  {title} {Arrays of individually controllable optical tweezers based on
  3d-printed microlens arrays},\ }\href {https://doi.org/10.1364/oe.386243}
  {\bibfield  {journal} {\bibinfo  {journal} {Opt. Express}\ }\textbf {\bibinfo
  {volume} {28}},\ \bibinfo {pages} {8640} (\bibinfo {year}
  {2020})}\BibitemShut {NoStop}%
\bibitem [{\citenamefont {Stuhlm\"{u}ller}\ \emph {et~al.}(2023)\citenamefont
  {Stuhlm\"{u}ller}, \citenamefont {Farrokhzad}, \citenamefont {Kuświk},
  \citenamefont {Stobiecki}, \citenamefont {Urbaniak}, \citenamefont
  {Akhundzada}, \citenamefont {Ehresmann}, \citenamefont {Fischer},\ and\
  \citenamefont {de~las Heras}}]{Stuhlmller2023}%
  \BibitemOpen
  \bibfield  {author} {\bibinfo {author} {\bibfnamefont {N.~C.~X.}\
  \bibnamefont {Stuhlm\"{u}ller}}, \bibinfo {author} {\bibfnamefont
  {F.}~\bibnamefont {Farrokhzad}}, \bibinfo {author} {\bibfnamefont
  {P.}~\bibnamefont {Kuświk}}, \bibinfo {author} {\bibfnamefont
  {F.}~\bibnamefont {Stobiecki}}, \bibinfo {author} {\bibfnamefont
  {M.}~\bibnamefont {Urbaniak}}, \bibinfo {author} {\bibfnamefont
  {S.}~\bibnamefont {Akhundzada}}, \bibinfo {author} {\bibfnamefont
  {A.}~\bibnamefont {Ehresmann}}, \bibinfo {author} {\bibfnamefont {T.~M.}\
  \bibnamefont {Fischer}},\ and\ \bibinfo {author} {\bibfnamefont
  {D.}~\bibnamefont {de~las Heras}},\ }\bibfield  {title} {\bibinfo {title}
  {Simultaneous and independent topological control of identical microparticles
  in non-periodic energy landscapes},\ }\href
  {https://doi.org/10.1038/s41467-023-43390-0} {\bibfield  {journal} {\bibinfo
  {journal} {Nat. Commun.}\ }\textbf {\bibinfo {volume} {14}},\ \bibinfo
  {pages} {7517} (\bibinfo {year} {2023})}\BibitemShut {NoStop}%
\bibitem [{\citenamefont {Morin}\ \emph {et~al.}(2016)\citenamefont {Morin},
  \citenamefont {Desreumaux}, \citenamefont {Caussin},\ and\ \citenamefont
  {Bartolo}}]{Morin2016}%
  \BibitemOpen
  \bibfield  {author} {\bibinfo {author} {\bibfnamefont {A.}~\bibnamefont
  {Morin}}, \bibinfo {author} {\bibfnamefont {N.}~\bibnamefont {Desreumaux}},
  \bibinfo {author} {\bibfnamefont {J.-B.}\ \bibnamefont {Caussin}},\ and\
  \bibinfo {author} {\bibfnamefont {D.}~\bibnamefont {Bartolo}},\ }\bibfield
  {title} {\bibinfo {title} {Distortion and destruction of colloidal flocks in
  disordered environments},\ }\href {https://doi.org/10.1038/nphys3903}
  {\bibfield  {journal} {\bibinfo  {journal} {Nat. Phys.}\ }\textbf {\bibinfo
  {volume} {13}},\ \bibinfo {pages} {63–67} (\bibinfo {year}
  {2016})}\BibitemShut {NoStop}%
\bibitem [{\citenamefont {Dong}\ \emph {et~al.}(2022)\citenamefont {Dong},
  \citenamefont {Turci}, \citenamefont {Jack}, \citenamefont {Faers},\ and\
  \citenamefont {Royall}}]{Dong2022}%
  \BibitemOpen
  \bibfield  {author} {\bibinfo {author} {\bibfnamefont {J.}~\bibnamefont
  {Dong}}, \bibinfo {author} {\bibfnamefont {F.}~\bibnamefont {Turci}},
  \bibinfo {author} {\bibfnamefont {R.~L.}\ \bibnamefont {Jack}}, \bibinfo
  {author} {\bibfnamefont {M.~A.}\ \bibnamefont {Faers}},\ and\ \bibinfo
  {author} {\bibfnamefont {C.~P.}\ \bibnamefont {Royall}},\ }\bibfield  {title}
  {\bibinfo {title} {Direct imaging of contacts and forces in colloidal gels},\
  }\href {https://doi.org/10.1063/5.0089276} {\bibfield  {journal} {\bibinfo
  {journal} {J. Chem. Phys.}\ }\textbf {\bibinfo {volume} {156}},\ \bibinfo
  {pages} {214907} (\bibinfo {year} {2022})}\BibitemShut {NoStop}%
\bibitem [{\citenamefont {Gnesotto}\ \emph {et~al.}(2020)\citenamefont
  {Gnesotto}, \citenamefont {Gradziuk}, \citenamefont {Ronceray},\ and\
  \citenamefont {Broedersz}}]{Gnesotto2020}%
  \BibitemOpen
  \bibfield  {author} {\bibinfo {author} {\bibfnamefont {F.~S.}\ \bibnamefont
  {Gnesotto}}, \bibinfo {author} {\bibfnamefont {G.}~\bibnamefont {Gradziuk}},
  \bibinfo {author} {\bibfnamefont {P.}~\bibnamefont {Ronceray}},\ and\
  \bibinfo {author} {\bibfnamefont {C.~P.}\ \bibnamefont {Broedersz}},\
  }\bibfield  {title} {\bibinfo {title} {Learning the non-equilibrium dynamics
  of {B}rownian movies},\ }\href {https://doi.org/10.1038/s41467-020-18796-9}
  {\bibfield  {journal} {\bibinfo  {journal} {Nat. Commun.}\ }\textbf {\bibinfo
  {volume} {11}},\ \bibinfo {pages} {5378} (\bibinfo {year}
  {2020})}\BibitemShut {NoStop}%
\bibitem [{\citenamefont {Chollet}(2022)}]{Chollet2022}%
  \BibitemOpen
  \bibfield  {author} {\bibinfo {author} {\bibfnamefont {F.}~\bibnamefont
  {Chollet}},\ }\href@noop {} {\emph {\bibinfo {title} {Deep learning with
  python}}}\ (\bibinfo  {publisher} {Manning Publications},\ \bibinfo {address}
  {New York, NY},\ \bibinfo {year} {2022})\BibitemShut {NoStop}%
\end{thebibliography}
\end{document}